\documentclass[epjST]{svjour}
\usepackage{amsmath,amssymb}
\usepackage{txfonts} 
\usepackage{graphicx}
\usepackage{changes}
\begin{document}
\title{Pattern dark matter and galaxy scaling relations.}
\author{Shankar C. Venkataramani\inst{1} \and Alan C. Newell\inst{1}}
\institute{                    
  \inst{1} Department of Mathematics, University of Arizona, 617 N. Santa Rita Ave., Tucson, AZ 85721, USA
 }

\abstract{
We argue that a natural explanation for a variety of robust galaxy scaling relations comes from  the perspective of pattern formation and self-organization as a result of symmetry breaking. We propose a simple Lagrangian model that combines a conventional model for normal matter in a galaxy with a conventional model for stripe pattern formation in systems that break continuous translation invariance. We show that the energy stored in the pattern field acts as an effective dark matter. Our theory reproduces the gross features of elliptic galaxies as well as disk galaxies (HSB and LSB) including their detailed rotation curves, the radial acceleration relation (RAR), and the Freeman law. We investigate the stability of disk galaxies in the context of our model and obtain scaling relations for the central dispersion for elliptical galaxies. A natural interpretation of our results is that (1) `dark matter' is potentially a collective, emergent phenomenon and not necessarily an as yet undiscovered particle, and (2) MOND is an effective theory for the description of a self-organized complex system rather than a fundamental description of nature that modifies Newton's second law.
} %end of abstract

\maketitle

\setcounter{tocdepth}{2}
\tableofcontents

\section{Introduction}
\label{intro}

%%%%%%%%%%%%%%%%%%%%%%%%%%%%%%%%%%%%%%%%%%%%%%%%
%
%		Introduction
%
%%%%%%%%%%%%%%%%%%%%%%%%%%%%%%%%%%%%%%%%%%%%%%%%

Understanding what we call dark matter and dark energy are two of the great intellectual challenges of our time. Each is a place holder for current ignorance. Dark energy, responsible for approximately 70\% of all the mass/energy in the universe, is posited to explain the accelerating expansion of the universe and acts as if the space between galaxies and galaxy clusters where the density is diminishing is endowed by an increase in pressure, completely alien to common experience. The current paradigm in cosmology attempts to capture its effect through a cosmological constant.

Dark matter, on the other hand is invoked to explain observations which indicate that the gravitational forces binding stars into galaxies, and galaxies into clusters are significantly larger than what can be accounted for by the amount of baryonic matter in the universe \cite{Trimble1987existence}. In the early 1930s, Oort \cite{Oort1932force} analyzed stars in the solar neighborhood and concluded that visible stars only accounted for about half of the mass needed to explain their observed vertical displacements from the galactic plane. Contemporaneously, Zwicky \cite{Zwicky1933Rotversciebung} studied the velocity dispersion of nebulae in the Coma cluster and deduced that the cluster required more than 100 times the mass in the luminous galaxies in order to stay bound. This led him to propose that the gravity of unseen {\em dunkle materie} (dark matter) was holding the cluster together.  The hunt for invisible matter became a serious endeavor in the wake of the pioneering findings of Rubin, Ford and Thonnard~\cite{Rubin_Rotation_1970,Rubin_Extended_1978} on the rotation curves of galaxies.  

\begin{figure}[htbp] %  figure placement: here, top, bottom, or page
   \centering
   \includegraphics[width=0.8\textwidth]{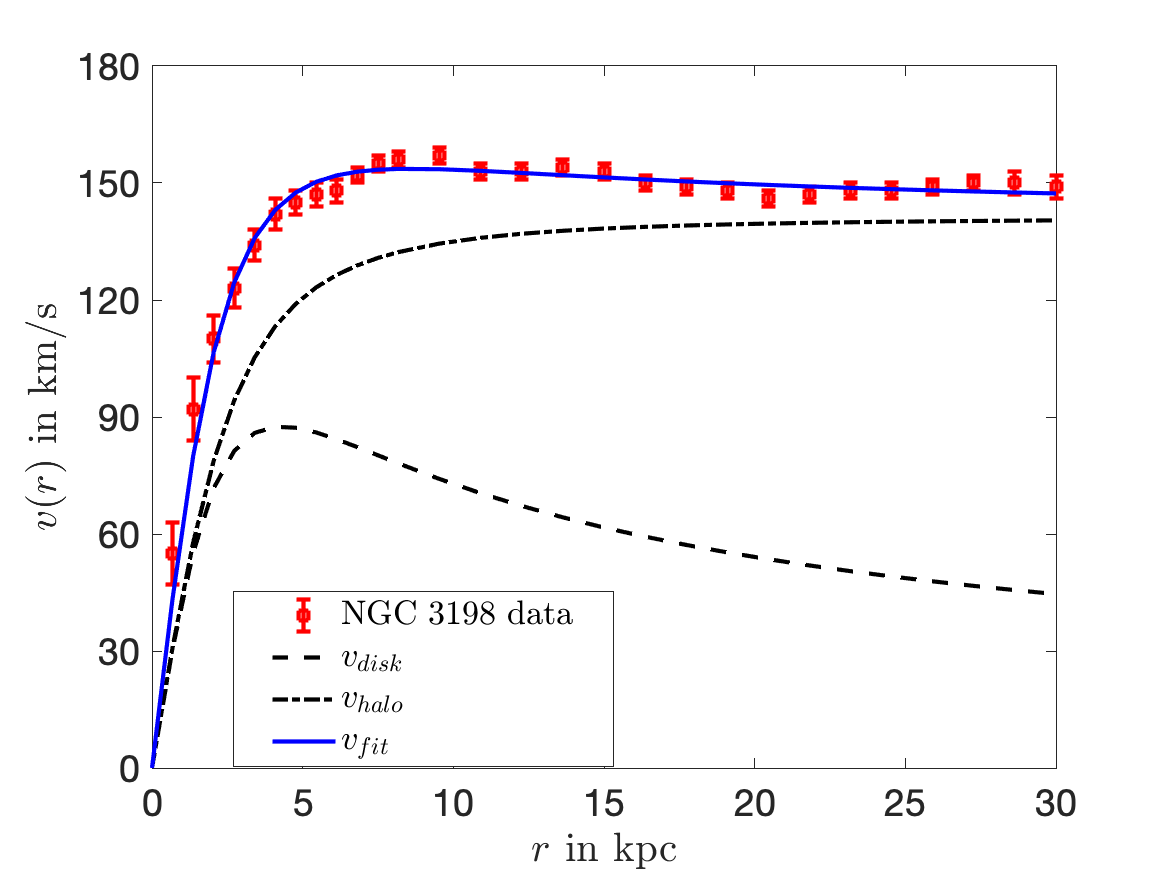} 
   \caption{The rotation curve for NGC 3198. $r$ is the distance from the galactic center and $v(r)$ is the rotation speed. The data is from van Albada {\em et al} \protect{\cite{van_Albada_1985}}. }
   \label{fig:rotation}
\end{figure}

A simple model in which the force experienced by the star executing a circular orbit at radius $r$ about the galactic center due to some large central mass $M$ leads to a Keplerian rotation curve $v = \sqrt{\frac{GM}{r}}$. Observations, however, consistently demonstrate that the galactic rotation curves flatten out and the orbital velocity of distant stars is roughly constant ~\cite{Rubin_Rotation_1970,Rubin_Extended_1978}. This is illustrated, for example, in fig~\ref{fig:rotation}  generated from rotation velocity data for the galaxy NGC 3198 (van Albada {\em et al} \cite{van_Albada_1985}). We can model the data by a  parameteric fit  of the form $v^2 = v^2_{\infty} \xi^2 \left((1+\xi^2)^{-1} + \alpha (1+\xi^2)^{-3/2}\right)$ with $\xi = r/r_0$, corresponding to a Kuzmin disk with a quasi-isothermal halo. The fitting procedure is phenomenological and not reflective of any underlying physical processes. Rather, it is purely for the purposes of describing the rotation curve. The success of the fitting procedure demonstrates the adequacy of using a quasi-isothermal halo in describing the flattening of the rotation curve of galaxies. 

Turning the argument around by balancing $GM/r^2$ with $v^2/r$ with $v$ constant, gives a mass $M$ of $v^2 r/G$ which is more mass than the galaxy would seem to contain. Dark matter (DM), which constitutes about 5/6 of all matter, and about five times the amount of baryonic matter (stars, gas, walls, you, me) observed, was invented to resolve this and other discrepancies  \cite{Zwicky_Masses_1937,Rubin_Rotational_1980}. For example, the data in Fig.~\ref{fig:rotation} by some invisible halo of matter of unknown origin surrounding them \cite{van_Albada_1985}. But of what ingredients is the new matter composed? The short answer to date is that nobody knows and the source of this extra mass has not yet been definitively identified. Also, there are countervailing view points. Milgrom \cite{Milgrom_MOND_1983} and others have argued that the observations should be interpreted as the need to modify Newton's second law at small accelerations (MOND), and that the dynamics on galactic scales can be explained  without any appeal to missing mass. 

The dominant paradigm in cosmology is the LCDM model which postulates that dark energy is modeled by a cosmological constant, and the bulk of matter in the universe consists of dynamically cold dark matter that has clumped into halos. These halos serve as the seeding grounds for galaxies and galaxy clusters consisting of baryonic matter. The latter lose energy by radiation and tend to gather at the lowest energy basins of the halos. Interaction, probably tidal in nature, between neighboring halos and subhalos endows individual clumps with angular momentum which is distributed in equal amounts to both the dark and baryonic matter. 

While the model fits many disparate observations, there are significant questions the answers to which need to be understood before LCDM is declared an unqualified success. For example, if one moves the cosmological constant term $\Lambda g_{ij}$ to the right hand side of Einstein's equation and interprets it as a part of the stress-energy tensor, it implies that the pressure density relation is (in the simplified case of an almost flat universe) $p = -\rho$, a relation so completely counter to common experience that its true meaning would have to be uncovered before one can place great confidence in the theory. While there are instances in nonlinear physics, the best examples are in optics, for which the Hamiltonian is non positive definite and which can be interpreted to behave in ways similar to that of a fluid with negative pressures, there is no understanding to date as to why such a system has relevance in the cosmological context. 

Likewise, there are weaknesses with the notion of dark matter. While the CDM model works remarkably well on cluster and cosmological scales, no plausible candidate for a dark matter particle has yet been uncovered despite the fact that intense searches in those parameter ranges where one expects to find such a particle have so far not born fruit. In addition, there are tensions between CDM predictions and observations on galactic and smaller scales. N-body simulations give halos with the so called universal Navarro-Frenk-White (NFW) density profile \cite{NFW96},
\begin{equation*}
\rho_{\text{NFW}}(R) = \frac{\rho_0 }{ [(1+(R/R_S)^2)(R/R_s)]}
%\label{eq:NFW}
\end{equation*}
 where $R$ is the 3d radial coordinate. Observations, however, favor ``cored" halos, for instance the quasi-isothermal profile 
 \begin{equation*}
 \rho_{\text{qiso}} = \frac{\rho_0}{(1+R^2/R_C^2)}
 %\label{eq:Qiso}
\end{equation*} 
over the ``cuspy" NFW profile \cite{Li2020Comprehensive}.

For one class of galaxies that we consider in this paper, namely cold disk galaxies, observations suggest that the surface brightness decays exponentially in $r$, the two dimensional radial coordinate as measured from the galaxy center. Assuming that surface brightness is a proxy for surface mass, the distribution of baryons is given by a surface density $\Sigma(r)= \Sigma_0 \exp(-r/r_0)$, where  $\Sigma_0$  and $r_0$ are the central density and baryonic scale lengths of the galaxy respectively \cite{BT08}. And what is remarkable is this. There are tight correlations and scaling relations between the halo parameters $\rho_0$ and $R_c$ and the baryonic parameters  $\Sigma_0$  and $r_0$ \cite{Donato2004Cores,Donato2009constant}, %.  Let us emphasize this. The 
i.e. the parameters defining the dark matter halo are connected with the parameters defining the baryonic matter. 

 Further, the flattening of the velocity rotation curve to an asymptotic ($r$ large but not so large as to involve a neighboring galaxy or cluster) value $v_{\infty}$  is given by a relation in which $v_{\infty}$  depends on the baryonic mass $M_B$, the gravitational constant $G$ and a universal acceleration $a_0$; namely $v^4=\alpha G M_B a_0$ where $\alpha$ is an order one constant. This relation is known as the Baryonic Tully-Fisher Relation (BTFR) \cite{McGaugh2000BTFR}. According to the Newtonian balance, $v(r)$ should decay as $(GMB/r)^{1/2}$. An additional mass, dark matter, was invoked in order to explain why $v(r)$ tends to a constant $v_{\infty}$  at large $r$. But that value, $v_{\infty}$, depends in a universal way on the baryonic mass in the galaxy and implies the relations $M_{\text{halo}} \gtrsim M_B, M_{\text{halo}} \propto M_B^{1/2}$ and there is no obvious explanation for why this connection should hold \cite{Famaey2012MOND,McGaugh2012BTFR}. 

These correlations between the baryonic and halo parameters are not unique to cold disk galaxies. Any model which purports to explain galactic behavior by appealing to the notion of dark matter and an enveloping halo has to account for this remarkable observation that galaxies are  seemingly governed by a single dimensionless  parameter  \cite{Disney2008Galaxies}.

A general organizing principle that can account for many of these correlations is the radial acceleration relation (RAR) which is a local relation between for the observed acceleration $g_\text{obs}$ and the purely baryonic acceleration $g_\text{bar}$ \cite{Lelli2017onelaw}. The RAR holds for a range of galaxies including dSphs, disk galaxies (SO to dIrr) and giant ellipticals. It was first proposed as the basis for MOND, modification of Newtonian dynamics. MOND was originally proposed by Milgrom and posits that at very low accelerations ($\sim 10^{-10}$ms$^{-2}$) The Newtonian acceleration $g_\text{bar} =GM/r^2$ is replaced by its geometric mean with a universal acceleration $a_0$; $g_\text{MOND}=\sqrt{g_\text{bar} a_0}$. MOND, although without a physical foundation, has proven to be remarkably valuable in that it has both predicted new results and been consistent with known observations. We note for example that the balance $v^2/r=g_\text{MOND}$ leads directly to BTFR. The success of MOND, admittedly a hypothesis without any obvious physical justification in classical physics, requires all theories that purport to explain observed behavior, including LCDM, to account for the fact that the observed acceleration seems to behave as if it follows MOND. Equally important, and as emphasized above, all such theories have also to explain how almost all observed quantities depend on the baryonic mass distribution which is mainly supported in the neighborhood of the galactic plane and not on the dark matter halo. In addition, a successful theory has to account in rotation dominated galaxies for a natural upper bound to the disk mass density known as the Freeman limit. Thus, despite the fact that galaxy formation is inherently stochastic, the BTFR, RAR, and the existence of the Freeman limit all suggest that some self-organizing principles may be at work \cite{Aschwanden2018order,plb}. 

Our aim in this paper is to suggest that indeed self-organizing mechanisms are playing a central role in galaxy formation and, to this end, we present a new perspective and a completely new theory that captures many of the scaling behaviors of galaxies. For rotation supported galaxies we can describe not only the flattening, but also all details of the velocity rotation curve. Moreover, we also find relations which parallel both BTFR and RAR and, remarkably, find that a relation derived from the energy of defects in patterned structures gives rise to the Freeman limit. For pressure supported systems we recover the Faber-Jackson relation \cite{Faber1976Velocity} and gain some insight into the mechanisms leading to the fundamental plane \cite{Djorgovski1987Fundamental,Gudehus1991Systematic}. Finally, we are can also characterize ``mixed" galaxies consisting of two components -- a pressure supported bulge coexisting with a cold rotation supported disk. 

 We are motivated in this endeavor by the general properties of pattern forming systems and an appreciation that instability generated patterns do have a role to play in the formation and structures of galaxies. In particular, we argue that patterned systems can store energy, energy that can lead to additional forces that can act in a way that gives rise to behaviors generally attributed to the presence of dark matter halos. Let us emphasize this. Energy that depends on the parameters describing baryonic matter and that arises from defect structures gives rise to forces that produce the observed effects attributed to dark matter. Thus it is very natural and not at all surprising that there should be tight correlations and scaling relations between halo and baryonic parameters.

Philosophically our approach is guided by the principle that it is wise to explore classical, perhaps subtle and non-obvious, explanations for phenomena before inventing completely new physics. In this quest, we are inspired by the pioneering work and ideas of Yves Couder \cite{Couder2006Single} who, together with colleagues and notably  John Bush \cite{Bush2010Quantum,Molacek2013Drops}, demonstrated how a classical system, obeying purely Newtonian laws and near the phase transition in a Faraday dish, could exhibit many of the mysterious behaviors associated with quantum physics provoking the question: Might the deterministic but chaotic dynamics of a classical system, here a resonant interaction between a bouncing particle and a companion pilot wave carrier, underlie quantum statistics? Whereas no one doubts that quantum mechanical thinking has had many extraordinary successes (think the success of Dirac's equation and virtual particles) in predicting atomic parameters to within one part in a billion, it is nevertheless sensible to apply Occam's razor and to ask if more prosaic interpretations could reproduce equivalent results. And, with his pioneering experiment of the curious behaviors he noticed with a bouncing droplet in a Faraday dish, Yves Couder had the imagination and the tenacity to do just that. Unfortunately the scientific community lost a great scientist when Yves died this past year. He was truly an original thinker and his ideas will have an impact on scientists for many generations to come. We honor his memory by taking a parallel path to his in our search for more prosaic explanations of ``dark matter".

In keeping with this philosophy, and with the appreciation that no broadly accepted candidate for dark matter has yet been identified, we will demonstrate that there are subtle behaviors associated with classical self-organizing systems that may provide possible alternatives to postulating new forms of matter. 

%%%%%%%%%%%%%%%%%%%%%%%%%%%%%%%%%%%%%%%%%%%%%%%%
%
%		A Roadmap
%
%%%%%%%%%%%%%%%%%%%%%%%%%%%%%%%%%%%%%%%%%%%%%%%%

\subsection{A roadmap} \label{sec:roadmap}

We begin in section~\ref{sec:phase_surfaces} by outlining the results from patterns that give the basis for our new approach to the role of self-organization in the dynamics of galaxies. Two key realizations are that patterns are macroscopic objects with universal descriptions that average over details of the microscopic origins and that topological and boundary constraints can lead to patterns with defects that do not access minimum energy configurations. The energy stored in such patterns has consequences and gives rise to what we call pattern dark matter in which this energy provides forces that can lead to behaviors associated with dark matter halos. Indeed, we will show that the corresponding equivalent mass density has precisely the shape of a cored halo. 

Following this, and motivated by the universal structure of the pattern average energy which we think of as playing the role of an action, we introduce in section~\ref{(3+1)stripes} an additional Lagrangian to will later be added standard Einstein-Hilbert action of GR. Our first goal is to show that such an action, which captures the energy associated with a pattern defect structure generated by the gravitational instability of a baryonic mass cloud, can produce an additional effective force that leads to a flattening of the velocity rotation curve. We title this section ``The origin of a pattern dark halos" because we will show that the energy associated with the defect structure is equivalent to a ``dark halo"  with mass $M_P \propto R$ which grows with the radius $R$. The resulting Newtonian acceleration, $GM_P/R^2$, then leads to a balance with the centrifugal acceleration with a constant rotational velocity $v_\infty$. 

To illustrate the main ideas, we begin with what is admittedly an unrealistic gravitationally induced structure, a spherical target pattern although many of the details introduced here will also be relevant when we update the calculation using a more realistic galactic model. In particular, it clearly illustrates the notion that defect structures in instability generated patterns behave as additional mass halos. 
In section~\ref{sec:Pattern_Halos},  we calculate the effective mass structure and the asymptotic behavior $v_{\infty}$ of the velocity rotation curve for such a spherical pattern halo.  

In section~\ref{sec:action}, we introduce the full Lagrangian action including the terms corresponding to the pattern action and an interaction term that couples the baryonic density to the pattern phase field. At this stage, our effective Lagrangian has two additional parameters $\Sigma^*$, a surface density which enters the dimensional factor multiplying the energy, and $k_0$, the pattern preferred wavenumber. The next challenge then is to interpret these parameters in terms of galactic parameters and in particular the total baryonic mass $M_B$ although we emphasize that, irrespective of these choices, the theory leads to behaviors, the flattening of the velocity rotation curve, the Tully-Fisher relation, the Freeman limit and the radial acceleration relation, generally associated with the influence of dark matter halos. While this is promising, it is very important that we make the link between the parameters in our action, $\Sigma^*$ and $k_0$ and  the parameters such as the universal acceleration scale $a_0$ and the total baryonic mass $M_B$ that appear, from many, many observations, to determine the dynamical behaviors of  galaxies.

This connection is made in section~\ref{sec:stability} where we combine the expression for $v_\infty$ which is given in terms of the pattern parameters $\Sigma^*$ and $k_0$ with two other relations arising from a stability analysis of a differentially rotating disk. Those two relations are the expression for the local preferred wavenumber $k_1(r)$ in terms of the disk surface density and a saturated Toomre parameter. The Toomre parameter $Q$ expresses the ratio of rotation stabilizing to gravitationally destabilizing influences in a rotating disk and we assume that an initially linearly unstable disk will saturate due to nonlinear feedback to give an effective $Q$ of unity. These three relations lead to an identification $\Sigma^*= \frac{a_0}{2\pi G}$ where $a_0 \simeq 10^{-10}$m/s\textsuperscript{2} is a universal acceleration and $k_0 \propto \sqrt{\frac{\Sigma^*}{M_B}}$. This latter identification is consistent with what we obtain using an entirely different approach in section~\ref{sec:dynamical}.  

We demonstrate in the sections that follow that our models give realistic velocity rotation curves, are recover various galaxy scaling relations including the baryonic Tully-Fisher relation (BTFR) and the existence of the Freeman limit for disk galaxies, the Faber-Jackson relation for elliptic galaxies, and the radial acceleration relation (RAR) for both disk and elliptical galaxies.

Section~\ref{sec:LTG} outlines the details, in three steps, of how the stationary points of the total action are calculated. The result for a spherical halo is repeated as an example. In section~\ref{sec:eikonal}, we apply this procedure to more realistic disk galaxy models. These model galaxies are axisymmetric and their masses are concentrated on the galactic plane. Nevertheless, they have halos, manifested as nontrivial structures in the phase field, with the phase contours essentially given by an application of Huygen's principle. The phase fronts are (approximately) spherical caps but are not anymore given by a spherical target. The contours intersect the galactic plane at an angle not equal to $\pi/2$ and this means that there is a discontinuity of the phase gradient on the galactic plane. Such structures are well known in pattern theory and their regularized shapes are called phase grain boundaries (PGB) with energy densities proportional to the cube of the sine of the angle $\theta(s)$ at which the contours intersect the galactic plane where $s$ measures the radial distance from the galactic center. The extrema of the action corresponding to this disk structure gives rise to a relation between the (baryonic) matter surface density $\Sigma_B(s)$, the angle $\theta(s)$ and the parameter $\Sigma^*$ in the action, a relation which turns out to imply the Freeman limit.

A remarkable consequence from our model is that the sine cubed dependence of the energy density, a result arising from pattern theory independent of its connection with galactic halos, naturally leads to matter distributions corresponding to Kuzmin disks. As we shall discuss, the Kuzmin disk plays a special role in the analysis of disk galaxies, akin to an attractor for dynamical systems. In section~\ref{sec:RC_disks} we derive in detail the rotation curves for Kuzmin and exponential disks for galaxies which are dynamically cold, meaning that they are purely rotation dominated with no significant random motions or three dimensional structure. We also demonstrate the existence of a radial acceleration relation (RAR) for such models. The analysis in sec.~\ref{sec:LTG} is therefore applicable principally to cold LSB galaxies with a prescribed (static) surface density/brightness that is everywhere below the Freeman limit.

In sections~\ref{sec:spherical} and \ref{sec:HSB} we investigate dynamical self-organization in our model to obtain self-consistent solutions through the appropriate galaxy distribution functions. We consider system with no a priori restrictions on the matter distribution or the velocities, so a measure of the efficacy of our modeling is its ability to model spherical (elliptic) as well as disk+bulge galaxies, and successfully recover various galaxy scaling relations including the fundamental plane for elliptic galaxies (sec.~\ref{sec:FP}) and the RAR for HSB galaxies (\ref{sec:RAR}). We conclude in section~\ref{sec:discuss} with a short discussion of the key features of our modeling framework, its successes as well as the outstanding challenges, and avenues for future exploration.

%%%%%%%%%%%%%%%%%%%%%%%%%%%%%%%%%%%%%%%%%%%%%%%%
%
%		Phase surfaces
%
%%%%%%%%%%%%%%%%%%%%%%%%%%%%%%%%%%%%%%%%%%%%%%%%

\section{Stripe patterns and dark matter halos} \label{sec:phase_surfaces}

\begin{figure}[htbp] %  figure placement: here, top, bottom, or page
\centering
   \includegraphics[height=0.5 \textwidth]{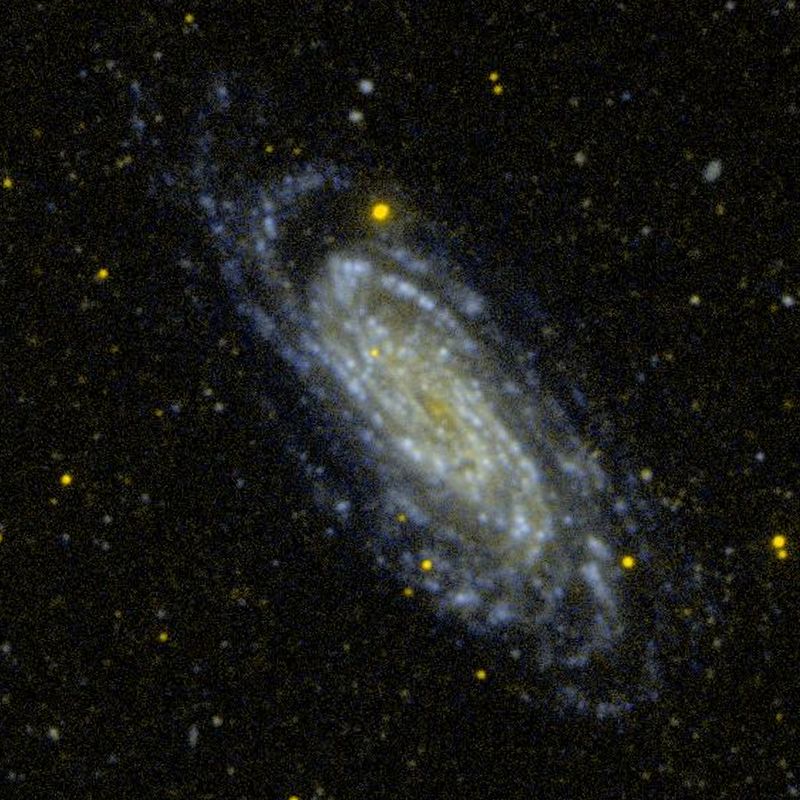} 
   \caption{The spiral galaxy NGC 3198 (imaged by GALEX).}
   \label{fig:spirals}
\end{figure}

The ideas we put forward are motivated by what happens in pattern forming systems. One only has to look around in nature to see how ubiquitous patterns are in nature; the ripples on long sandy beaches and windswept dunes, cloud formations, the growth stems of many plants such as sunflowers, the skin of cacti, the elastic sheet type wrinkles in many plant leaves, and even in highly turbulent situations such as the sun's surface there appear relatively ordered buoyancy driven cellular and granular structures resulting from gravitational instabilities. Patterns occur on a range of length scales, from viral capsids to fingerprints to the spiral arms of galaxies (see Fig.~\ref{fig:spirals}). Such patterns arise when systems are driven far from equilibrium by some external stress which stress, when crossing some critical threshold, leads to the destabilization of one state and the emergence of another. At the phase transition, some symmetries are broken and this leads to the preferential amplification of certain shapes and configurations. These amplified modes, and in some cases Goldstone modes which are neutral at the phase transition also play a role, compete via nonlinear interactions and a winning configuration, or sometimes an equivalence class of winning configurations, depending on the role played by some symmetries unbroken at the transition, emerge. 

The granddaddy of pattern forming systems is convection in a horizontal layer of fluid heated from below and we find it useful to explain some of our ideas using this example. At small adverse temperature gradients, heat is carried across the layer by conduction. At a certain temperature difference threshold, however, the conduction state becomes unstable and leads, when the layer is symmetric about its mid-plane, to a convection state of roll like motions in which the continuous translation symmetry is broken and is replaced by a discrete translation symmetry reflecting the choice of a preferred roll wavelength. Whereas translational symmetry is broken, rotational symmetry is not and so, in the emerging state of convection, the direction of the roll axes is chosen by local biases such as boundaries, imperfections or other constraints such as rotation. For fluids with moderate to low Prandtl numbers, certain neutral soft modes driven by pressure differences resulting from differing local intensities in convection also play a role but, for simplicity, we will here concentrate on the simpler case typified by high Prandtl number convection where such modes are absent. The consequences of the broken translational and unbroken rotational symmetry are very important because, in large (aspect ratio) systems where the box size is much larger than the preferred wavelength, the resulting state is not the lowest energy state whereby all rolls point in the same direction. Rather, the emerging pattern is a mosaic of patches of almost constant direction that meet and meld along boundaries, lines and points in two dimensions, planes, lines and points in three, at which the wave-vectors change fairly abruptly. The resulting defects play hugely important roles in large patterned systems as they carry both topological charges and they carry energy. The topological charges of circulation and twist, measuring concentrations of ``vorticity" and Gaussian curvature of an associated phase surface, reflect constraints imposed by distant boundaries and in many cases these constraints mean that the pattern can never reach its absolute minimum energy state. As a consequence, these metastable states with defects carry energy, significant amounts of energy. We use the term metastable to connote that they are local energy minima if constraints dictate they are minima in the restricted function spaces, or that, because coarsening processes take so long, they are, to all extents and purposes, local minima in that their unstable manifolds are extremely weak. The upshot is that patterns with defects contain energy and that that energy has consequences. We shall argue that energy stored in defects gives rise to additional forces that cause the flattening of the velocity rotation curves in galaxies and to the many observed scaling properties. The dominant defect containing patterns in rotation dominated disk galaxies have the structures of targets and spirals.

A powerful idea, that goes to back to the work of Landau \cite{Landau1937theory}, is the use of  {\em universal order parameter equations} to study phase transitions. An order parameter is a quantity that reflects the particular symmetries that are broken in the phase transition, is (typically) zero in the homogeneous state, but takes on a non-zero value in the patterned state. The order parameter therefore reflects the particular symmetries that are broken (and thus also the symmetries which are not broken) in the phase transition. They have two important properties. First, order parameters are the active mode coordinates that emerge from the unstable and neutral manifolds of the unstable state. The coordinates of the passive modes, those associated with the stable manifold, are slaved by algebraic expressions to the coordinates of the active modes. Second, the order parameters satisfy universal equations which depend principally on the symmetries of the underlying microscopic system they are analyzing and are insensitive to the precise details of that underlying system. For systems that support striped patterns but for which rotational symmetry remains unbroken at the phase transition and which do not have soft modes, the appropriate order parameter is the phase $\psi$ of the locally periodic structure and its gradient  $\mathbf{k}$ which gives the local orientation of the pattern. Examples include high Prandtl number convection and wrinkles on elastic sheets. In the former case, there is an additional feature that can add a richer structure to the topological indices of defects (they can be half integer and even third integer valued) and that feature is that the so called wave-vector $\mathbf{k}$ is not a vector field but rather a director field. But that feature will not play a significant role in what we discuss here as the defects we consider are mainly targets and spirals for which the twist indices are integer valued. It is their stored energies that are significant. 

In these systems, the order parameter evolution is essentially a gradient flow in which the energy $E$ is an average of the underlying microscopic energy functional over the local wavelength of the pattern \cite{passot1994towards,NV17}. It takes on a canonical form,
\begin{equation}
E = \frac{E_0}{k_0^2} \int  \left[\left(|\nabla\psi|^2 -k_0^2 \right)^2 + \left(\Delta\,\psi\right)^2\right]  dxdy, 
\label{CNred}
\end{equation}
where $\psi$ is non-dimensional, $E_0$ is an energy scale needed for dimensional consistency, lengths are nondimensionalized by the preferred wavenumber $k_0$ and the relevant nondimensional parameter is an aspect ratio is $(k_0 L)^{-1}$ where $L$ is the macroscopic scale of the pattern, i.e. the typical distance between defects or the size of the domain. The effective energy~\eqref{CNred} gives a good description of stripe patterns in the limit $k_0^{-1} \ll L$.
 
 The universal form of the average energy is interesting in that it is a combination of the first two differential forms of the phase surface $\psi=$ constant \cite{NV17,Newell2019pattern}. The first term, corresponding to a stretching energy in the elastic sheet context, depends on coordinate invariant combinations of the metric two form. The second term, corresponding to the bending energy in the elastic context, consists of coordinate invariant combinations of the curvature two form. Indeed in general contexts we often refer to these energy contributions as the stretching and bending energies. The curvature form takes on two parts. One is due to mean curvature and the other to Gaussian curvature. In two dimensions, it is for all practical purposes the determinant of the Hessian of the phase surface $\psi$; in three dimensions, the determinants of minors of the Hessian matrix. In all cases, these contributions can be converted to boundary integrals and measure important topological indices which, however, in the present context are important but will not be central to our story.

The ground or minimum energy state corresponds to parallel stripes with wavenumber $k_0$ for which $E=0$. If boundaries or other dynamic constraints such as angular momentum conservation dictate that the the pattern be radial, $\psi(\mathbf{x})=\psi(R)$ where $R$ is a radial coordinate, we cannot be in a ground state. Indeed, if we seek radial minimizers $\psi=\psi(R)$, we find $\mathbf{k}$ tends to zero as $R$ tends to zero and $\psi$ tends to $R$ for large $R$. These target patterns are robust because they cannot be continuously deformed into the plane wave ground states. Moreover, in large aspect ratio systems, because the local pattern is locally stripe like (the radii of curvatures of the targets are large compared to the pattern wavelength) we can represent the average energy of such patterns by~\eqref{CNred}. 

And the key observations now are these. The energy density of such patterns as function of $R$ is bounded for small $R$ and decays as $1/R^2$. Consequently if we were to think of the field $\psi$ as representing a dark matter spherical halo, we would note the following important features \cite{Newell2019pattern,plb}:
\begin{enumerate}
    \item  First, it produces a quasi-isothermal halo with $R_c=k_0^{-1}$. 
    \item Second, integrating over a volume of radius $R$ means that the accumulated energy grows linearly with increasing $R$. If we interpret this energy as an effective mass $M(R)$, then the effective mass of the phase field halo also grows as $R$. The Newtonian acceleration it will produce on a rotating star is $GM(R)/R^2$ and behaves as $1/R$ which, when balanced with the star's centrifugal acceleration, gives a rotation velocity $v(R)$ which is constant. In short, the pattern with a target defect gives rise to an additional force beyond the ordinary Newtonian force from a central mass and this force results in a different behavior of the rotation velocity. Indeed, we will shortly do this calculation and show that the velocity $v(R)$ tends to $v_{\infty}^2= \frac{16 \pi G\Sigma^*}{k_0}$. 
    \item Third, whereas we have chosen to do this calculation for a perfectly spherical field $\psi$, it is not hard to see that even if we had taken the phase surfaces of $\psi$ to be axisymmetric oblate spheroids rather than perfect spheres, the effects would be similar in that the mean curvature contribution to the bending energy would depend on $1/R^2$ where $R$ is the smaller of the two radii of curvature although there will be some compensation for the smaller volume element size. And indeed, when we calculate in sections 4, 5 the effects of the extra energy arising from the halos for a family of more realistic disk galaxies, including the Kuzmin and exponential disks, we obtain similar results. The stored pattern energy does indeed give rise to a force that, when balanced with the centrifugal force leads to a flattening of the velocity rotation curve.  As already noted, we will also find additional behaviors connected with the discontinuities on the galactic plane that connect the surface mass densities of the disks with the Freeman limit surface density.
\end{enumerate}

\subsection{Stripe patterns in spacetime: The origin of pattern dark halos} \label{(3+1)stripes}

A curved space-time generalization Cross-Newell energy \cite{newell1996defects} %that describes stripe patterns  
can be obtained from the `minimal coupling' assumption \cite{MTW} as  
\begin{equation}
\mathcal{S}_P := \int \mathcal{L}_P \sqrt{-g} d^4x = \frac{\Sigma^* c^2}{k_0^3}   \int \left\{(\nabla^\mu \psi \nabla_\mu \psi-k_0^2)^2 +  (\nabla^\mu \nabla_\mu \psi )^2\right\} \sqrt{-g} \,d^4x,
\label{RCN_4d}
\end{equation}
where $\mathcal{L}_P$ is the pattern Lagrangian density, $d^4x$ has dimensions of a length to the 4th power, the dimensionless metric $g_{\alpha \beta}$ has signature $(- \,+\, +\, +)$ and $\nabla_\mu$ is the corresponding covariant derivative.  Eq.~\eqref{Lagrangian} gives the {\em natural} covariant generalization \cite{NV17} of the {\em universal averaged energy} for nearly periodic stripe patterns \cite{passot1994towards}, 
and is thus expected to describe the macroscopic behavior of phase hyper-surfaces in curved spacetimes for a variety of microscopic models \cite{NV17}. The phase $\psi$ is dimensionless, $k_\mu := \nabla_\mu \psi$ has dimensions of inverse length, $\Sigma^*$ is a surface mass density scale, $k_0$ the preferred wavenumber, $c$ the velocity of light so that $\Sigma^*c^2/k_0^3$ is a normalizing constant to ensure that $\mathcal{S}_p$ has the correct dimensions for an action (the spacetime integral of an energy per unit volume).

The Euler-Lagrange equation for the Lagrangian in~\eqref{Lagrangian} is the 4th-order nonlinear wave equation
\begin{equation}
    \nabla^\alpha \nabla_\alpha \nabla^\beta \nabla_\beta \psi - 2 \nabla^\gamma \left\{( g^{\alpha \beta} \nabla_\alpha \psi \nabla_\beta \psi -k_0^2) \nabla_\gamma \psi\right\} = 0,
    \label{eq:full}
\end{equation}
on Minkowski spacetime with coordinate $(x^0= ct, x^1=x,x^2=y,x^3=z)$, (inverse) metric $\eta^{\alpha \beta}$ and signature $(- \,+\, +\, +)$. We seek stationary spherical target solutions $\psi = \psi(R)$ which both reflect the galactic halo and are ``localized", so that wavenumber mismatch $|\nabla \psi| - k_0$ vanishes as $r \to \infty$. 
For radial solutions $\psi = \psi(R) = \psi(\sqrt{x^2+y^2+z^2})$,  we have
\begin{equation}
\left(\partial_{RR} + \frac{2}{R} \partial_R\right)^2 \psi(R) - \frac{2}{R^2} \partial_R \left\{ R^2( \psi'(R)(\psi'(R)^2 - k_0^2) )\right\} = 0
\label{eleqn}
\end{equation}
While $\psi(R) = k_0 R$ is a solution of~\eqref{eleqn}, this solution is not smooth at the origin. Nonetheless, we expect this will describe the large $R$ behavior of $\psi(R)$, and $\psi'(R) \to 0$ as $R\to 0$ for the small $R$ behavior. We can rewrite~\eqref{eleqn} in flux-conservation form
\begin{equation}
\partial_R(R^2 \Gamma) = 0, \quad \Gamma = \partial_R \left[\frac{\partial_{R} (R^2\psi'(R))}{R^2}\right] - 2 \psi'(R)(\psi'(R)^2 - k_0^2) 
\label{flux}
\end{equation}
Setting $\psi(R) = k_0 R$ gives $\Gamma = -2 k_0/R^2 \neq 0$ although it is still true that $\partial_R (R^2 \Gamma) = 0$. Indeed, this is related to the lack of smoothness at $R=0$, as any smooth solution should satisfy $\Gamma = 0$ at $R=0$ and hence also everywhere. 

\begin{figure}[htbp] %  figure placement: here, top, bottom, or page
  \centering
  \includegraphics[width=0.8 \textwidth,trim={0cm, 1cm, 0cm, 2cm}, clip]{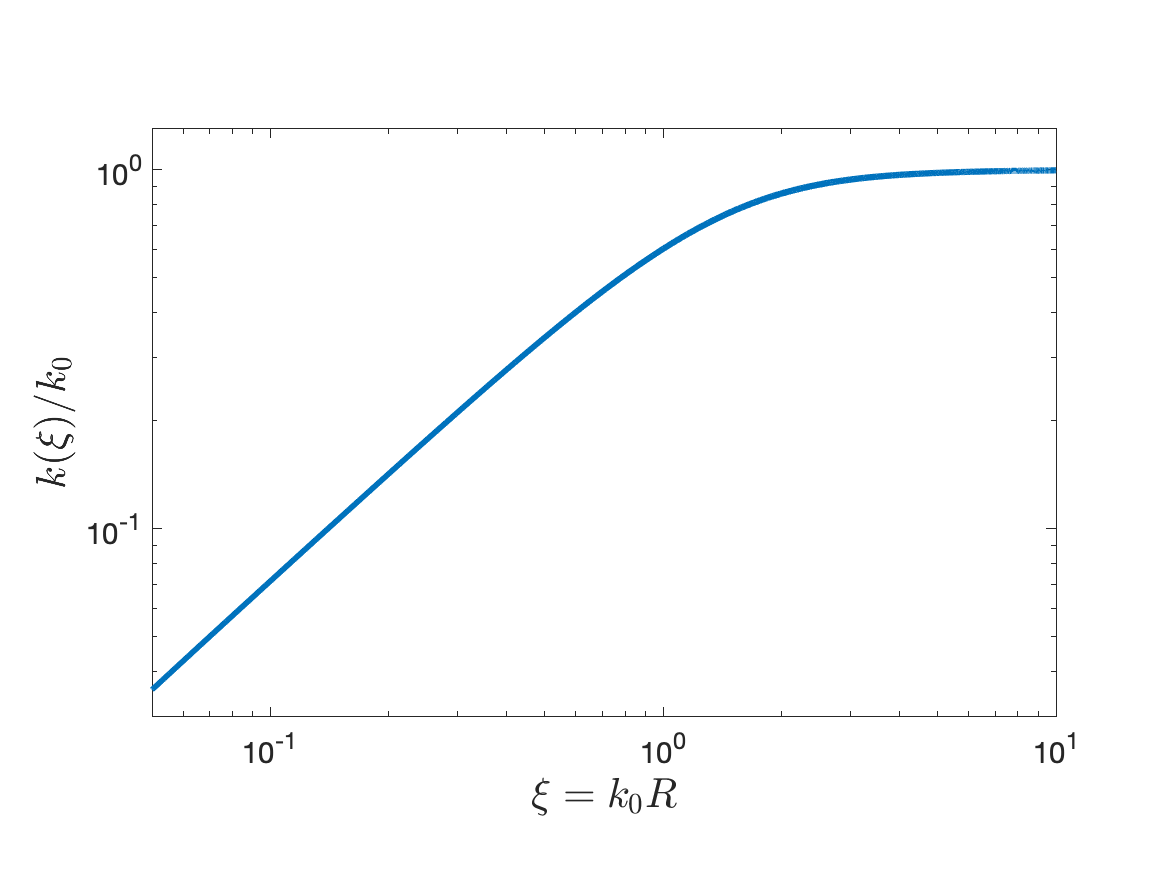} 
  \caption{The ``local" wavevector $k = \nabla \psi$ for a 3D target pattern obtained by numerically solving of the boundary value problem $k''(\xi) + 2 k'(\xi)/\xi -2 k(\xi)(k^2(\xi) + \xi^{-2}-1)=0, k(0) = 0$ and $k(\xi) \to 1$ as $\xi \to \infty$.}
  \label{fig:k-halo}
\end{figure}

The condition $\Gamma = 0$ is a {\em second-order} equation for $\psi'$ that we solve numerically using a spectral method \cite{Driscoll2014}.  Fig.~\ref{fig:k-halo} depicts the numerically obtained solution of the BVP $\Gamma =0$ using a shooting method to satisfy the boundary conditions $\psi'(0) = 0, \psi'(R) \to k_0$ as $R \to \infty$.

We get additional insight by analytically determining the far field behavior of the stationary target pattern. Linearizing about the leading-order far-field  solution, $\psi(R) = k_0 R + \alpha (k_0 R)^\beta$, we get power matching in the flux $\Gamma$ at leading order only for $\beta = -1$. For $\beta = -1$, we get
$$
\Gamma = \frac{(4 \alpha -2)k_0}{R^2} + O(R^{-4}) 
$$
   so that the far field corrections are given by $\beta = -1, \alpha = \frac{1}{2}$. This yields
   \begin{equation}
   \psi'(R) = k_0\left(1 - \frac{1}{2k_0^2 R^2} + O((k_0R)^{-4}) \right).
   \label{far-k}
   \end{equation}

\subsection{Pattern dark matter} \label{sec:Pattern_Halos}

The pattern Lagrangian~\eqref{RCN_4d} couples the pattern field $\psi$ to the geometry of the space time, and will thus contribute to the dynamically inferred ``gravitating mass".  We now outline an argument to compute the effective mass density corresponding to the pattern. 

 We  consider a static, spherically symmetric, curved space-time with the following weak-field metric given in {\em isotropic coordinates} \cite{MTW}, 
\begin{align}
\label{eq:metric}
g & = -(c^2 + 2 \phi(R)) dt^2 + (1 - 2 c^{-2} \phi(R)) (d R^2 + R^2 d\Omega^2). 
\end{align}
Let $\epsilon \ll 1$ denote the dimensionless small parameter, that governs the ``strength of gravity" i.e. the deviation in the metric from the flat Minkowski spacetime, so that $c^{-2} \phi$ is $O(\epsilon)$ and $\phi$ is the equivalent Newtonian potential. $\epsilon$ depends on the particulars of the system, and for example, for an isolated spherical mass $M$ of radius $a$, $\epsilon = \frac{GM}{c^2 a}$. 

Identifying our spacetime with  $\mathbb{R} \times \mathbb{R}^3$, the worldline of a nonrelativistic particle of rest mass $m_0$ is given by a vector $\mathbf{r}(t)$ in $\mathbb{R}^3$ with $\left|\frac{d \mathbf{r}}{d\tau}\right|^2 = \frac{1}{c^2}\left|\dot{\mathbf{r}}\right|^2 = O(\epsilon)$. Consequently, the action, the integral of the rest energy with respect to  proper time along the worldline, which is extremized for geodesics, is given by 
$$
-\int m_0 c \left[c^2 + \left(2  \phi(\mathbf{r}) - \left|\dot{\mathbf{r}}\right|^2\right)+O(\epsilon^2)\right]^{1/2} dt = \int \left[-m_0c^2 +  m_0\left(\frac{1}{2} \left|\dot{\mathbf{r}}\right|^2-\phi(\mathbf{r}) \right) + O(\epsilon^2)\right] dt
$$
Comparing with the classical action for a particle in a gravitational potential $\mathcal{S} = \int (T-V) dt$ we see that the (equivalent) Newtonian potential for a spacetime with the metric~\eqref{eq:metric} is given by the function $\phi = -\frac{1}{2}(c^2+g_{tt})$ provided that $|g_{tt}+c^2| \ll c^2$.

Retaining terms up to order $O(\epsilon)$, i.e. terms that are independent of or linear in $\lambda, \phi$, the pattern action $\mathcal{S}_P$ in~\eqref{RCN_4d}, for a radial pattern $\psi(R)$, is given by
\begin{align*}
\mathcal{S}_P  = & \frac{\Sigma^*c^2}{k_0^3} \int \left[\left(\psi'^2\left(1 - \frac{2  \phi}{c^2}\right)-k_0^2\right)^2+ \left(\left(\psi'' + \frac{2}{R} \psi'\right)\left(1-\frac{2\phi}{c^2}\right) \right)^2   \right]  \left(1 + \frac{2\phi}{c^2} \right)\, \times \nonumber \\
& R^2 dR \, d\Omega \,dt + O(\epsilon^2)
% \label{pattern_energy}
 \end{align*}
 
\begin{figure}[htbp] %  figure placement: here, top, bottom, or page
  \centering
  \includegraphics[width=0.8 \textwidth,trim={0cm, 1cm, 0cm, 2cm}, clip]{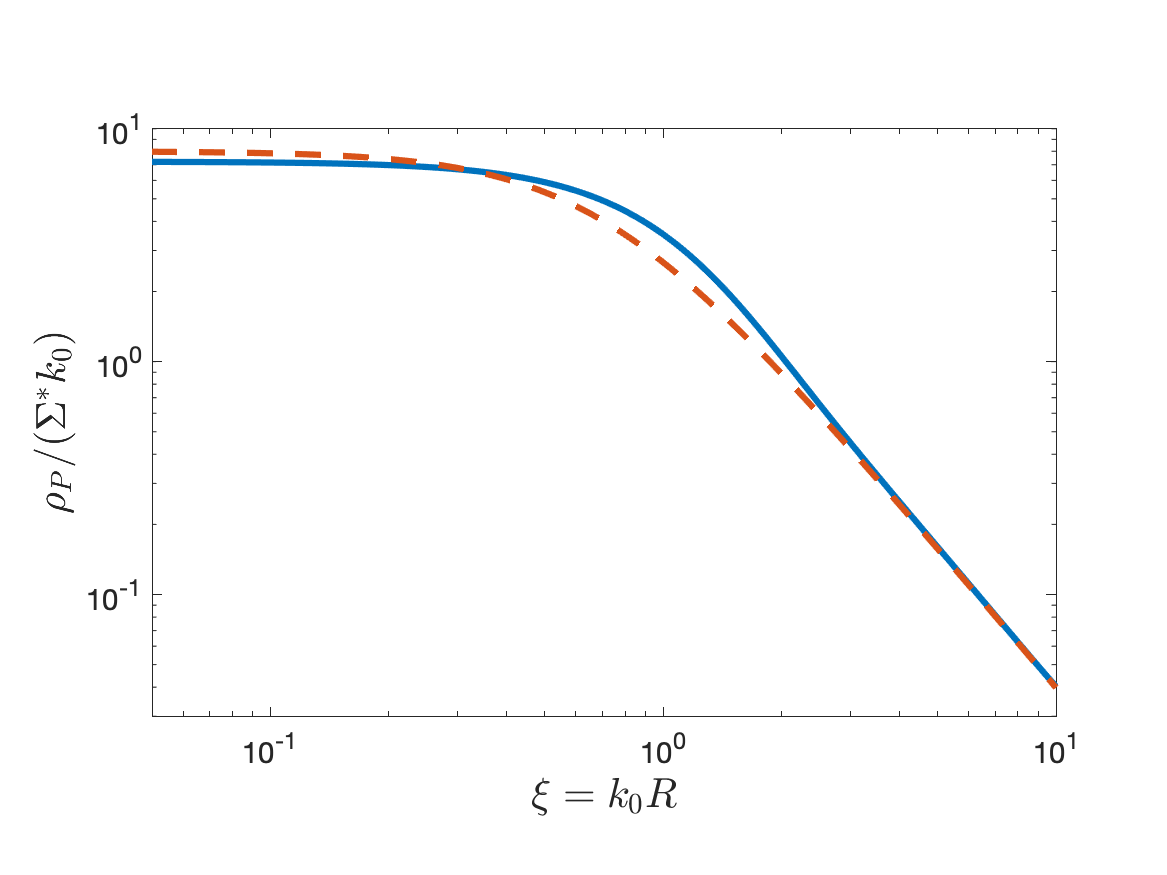} 
  \caption{The normalized effective pattern halo density obtained as a variational derivative (solid blue curve) and a rational approximation (dashed brown curve) given by $8/(1+2\xi^2)$.  }
  \label{fig:halo}
\end{figure}

We can determine the effective mass density in the pattern field by recognizing that, in the Newtonian limit, the density is given by the variational derivative $\rho =  \frac{\delta}{\delta \phi} \mathcal{V}$, where $\mathcal{V} = \int \rho \phi \,d^3x$ is the gravitational potential energy. Correspondingly, we obtain
\begin{align}
\rho_P & = - \frac{\delta}{\delta \phi} \mathcal{\mathcal{S}_P}= \frac{2\Sigma^*}{k_0^3} \left[\left(\psi'^4-k_0^4\right)+ \left(\psi'' + \frac{2}{R} \psi' \right)^2  \right] 
\label{halo-density}
\end{align}
Using~\eqref{far-k} we get the far-field behavior
\begin{equation*}
\rho_P = \frac{2\Sigma^*}{k_0^3} \left[\left(k_0^4 - 4 \cdot k_0^3 \cdot \frac{1}{2k_0 R^2} -k_0^4\right)+ \frac{4 k_0^2}{R^2}  \right]  + O(R^{-4}) \approx \frac{4 \Sigma^*}{k_0 R^2},
\end{equation*}
so that pattern DM halos have a ``universal" decay $\rho_P \approx 4 \Sigma^* k_0^{-1} R^{-2}$ in our theory, reflecting the far field behavior of the wavenumber of the target pattern. Fig.~\ref{fig:halo} is a plot of the effective halo density $\frac{\rho_{P}}{\Sigma^* k_0}$ given by \eqref{halo-density} as a function of the nondimensional radial coordinate $\xi = k_0 R$. We also plot, for comparison, a rational approximation $\frac{\rho_{P}}{\Sigma^* k_0} = \frac{8}{1+2\xi^2}$. 
The corresponding halo mass function is thus (approximately) given by  
\begin{equation}
M_P(R) =  4 \pi \int_0^R \rho_P(\zeta) \zeta^2 d\zeta   \approx \frac{8 \pi \Sigma^*}{k_0} \left(2 R - \frac{\sqrt{2}}{k_0} \arctan\left(\sqrt{2}k_0 R\right) \right).
\label{eq:halo_mass}
\end{equation}
corresponding to a quasi-isothermal halo, rather than the more commonly invoked ``cuspy" NFW halo. If the baryonic distribution is given by $M_B(R) = 4 \pi \int_0^R \rho_B(\zeta) \zeta^2 d\zeta$, equating the total gravitational acceleration $G(M_B(R)+M_P(R))/R^2$  with the centripetal acceleration $v^2/R$ of a circular orbit, yields
\begin{align}
\label{azimuthal}
v(R)^2 & \approx \frac{G M_B(R)}{R} + \frac{16 G \pi \Sigma^*}{k_0} \left(1- \frac{1}{\sqrt{2}k_0 R} \arctan\left(\sqrt{2} k_0 R\right) \right).
\end{align}
demonstrating the flattening of the rotation curve with an asymptotic velocity given by 
\begin{equation}
v_\infty^2 \approx 16 G \pi \Sigma^* k_0^{-1}.
\label{eq:asymptotic}
\end{equation}
Whereas this calculation provides encouragement that pattern structures with defects can provide additional forces resulting in behaviors similar to those for which the existence of dark matter was invoked, there are still challenges. Principal among these are: What determines the parameters $\Sigma^*$ and $k_0$ and how are they related to the distribution of baryonic matter? How do we eliminate the assumption of spherical symmetry and deal with models which are more realistic descriptions of disk galaxies? We discuss these issues in the sections that follow.

%%%%%%%%%%%%%%%%%%%%%%%%%%%%%%%%%%%%%%%%%%%%%%%%
%
%		The Action principle
%
%%%%%%%%%%%%%%%%%%%%%%%%%%%%%%%%%%%%%%%%%%%%%%%%

\section{The Lagrangian for pattern dark matter.} \label{sec:action}

We now introduce our action principle, that represent the influence of the stored energy in the pattern defect as well as the empirical observations/theoretical models that suggest a coupling between ``dark matter", here the pattern halo, and the baryonic density $\rho_B$ \cite{Sancisi2004visible,Famaey2020BIDM}. We posit a total action as the sum of
\begin{align}
\mathcal{S}_{EH} & = \frac{c^4}{16 \pi G}\int R \sqrt{-g} \, d^4 x, 
\quad \mathcal{S}_M   = \int \rho_B u^\alpha u_\alpha \sqrt{-g} \, d^4 x, \nonumber \\
-\mathcal{S}_P&  = -\frac{\Sigma^* c^2}{k_0^3}  \int \left\{(k_0^2-\nabla^\mu \psi \nabla_\mu \psi)^2 + (\nabla^\mu \nabla_\mu \psi )^2\right\} \sqrt{-g} \ d^4x, \nonumber \\
-{\mathcal S}_{\psi} & = -\int \rho_B c^2 V\left[k_0^{-2} g^{\alpha \beta}\nabla_\alpha \psi \nabla_\beta \psi \right] \sqrt{-g} \,d^4 x, 
\label{Lagrangian}
\end{align}

We have introduced an interaction term $S_{\psi}$ that couples the baryonic density $\rho_B$ and the pattern field $\psi$. The interaction is through the density $\rho_B V(|\mathbf{k}|^2)$ where $V(|\mathbf{k}|^2) \geq 0$ is a convex function of $\frac{\mathbf{k}}{k_0}$ that has a global minimum at $\mathbf{k} = 0$. Large values of $\rho_B$ creates ``defects" in $\psi$, like the spherical target pattern with $\nabla \psi = 0$ at the center. More generally, on a space-like slice of a (background) Minkowski spacetime, Jensen's inequality \cite{boyd2004convex} gives 
$$
\int \rho_B V(|\mathbf{k}|^2) d^3 x \geq M_B V(|\bar{\mathbf{k}}|^2)  
$$
where $M_B = \int \rho_B d^3 x$ and $\displaystyle{\bar{\mathbf{k}} = \frac{1}{M_B} \int  \rho_B \mathbf{k} d^3 x}$ is the mean wave-vector. In particular, the interaction term will try to achieve $\bar{\mathbf{k}}=0$ and will penalize the fluctuations $\mathbf{k} - \bar{\mathbf{k}}$.

$\mathcal{S}_P$ and $\mathcal{S}_\psi$ come with negative signs since they are `potential'  terms, i.e. akin to $W$  in the action $S = \int (T-W) \,dt$ for classical particle systems. 
What remains is the specification of $\Sigma^*, k_0$ and the potential $V$.
Before we do so, however, we emphasize that, irrespective of
the choices we are about to make, the additional terms $\mathcal{S}_p$ and
$\mathcal{S}_\psi$ in the action automatically lead to three of the key outcomes
that characterize the behaviors usually associated with
dark matter. First, we find that the curvature term in $\mathcal{S}_p$ leads to
an additional force which behaves as $1/r$
for large $r$ and a flattening
of the velocity rotation curve given by $v_{\infty}^2 \propto G \Sigma^*/k_0$.
Second,
the coupling between the pattern ``dark matter" and the baryonic
density leads to the Freeman limit with a maximum central
surface density on the universal scale, which we will shortly identify as being proportional to $\Sigma^*$ for rotation supported
systems. Third, the model predicts a radial acceleration relation
(RAR) between the total gravitational acceleration $g_\text{obs}$ and the
baryonic contribution $g_\text{bar}$, that closely resembles what is observed and, interestingly, has two branches.

We now turn to the identification of the choices for $\Sigma^*$ and $k_0$. There is substantial observational evidence for the existence of an universal acceleration scale $a_0$, or equivalently, a surface density scale $\Sigma^* = \frac{a_0}{2 \pi G}$ in the dynamics of galaxies. In earlier work \cite{plb}, we suggested that $k_0 \sim \sqrt{\Sigma^*/M_B}$, so that, it is not universal, but rather depends on the host galaxy in a ``nonlocal" manner through the total baryonic mass $M_B$.  This choice of $k_0$ recovers the Baryonic Tully-Fisher relation
$$
v_\infty^4 \propto \left(\frac{G \Sigma^*}{k_0}\right)^2 \sim G M_B a_0.
$$
It is therefore very desirable to identify a mechanism that ``dynamically" determines this value of $k_0$ for a galaxy, thus giving insight into the origin of the BTFR \cite{McGaugh2000BTFR}.

\subsection{The stability of rotation supported cold disks} \label{sec:stability}

Since our model is an attempt to encode the effects of instability induced patterns and self-organization, it is natural to consider the relation between the pattern field $\psi$ and the gravitational clumping instability in baryons. Indeed, the role of the dark matter halo in stabilizing the gravitational instabilities of a differentially rotating cold disk have already been identified in the work of Ostriker and Peebles \cite{Ostriker1973Numerical} and further investigated in the context of MOND \cite{Milgrom1989Stability,Brada1999Stability}. A WKB analysis the linear stability analysis for the differentially rotating gaseous disk gives 
\begin{equation}
(\omega-m \Omega)^2 = \sigma^2 \left(k-\frac{\pi G \Sigma_B}{\sigma^2}\right)^2 + \kappa^2 - \left(\frac{\pi G \Sigma_B}{\sigma}\right)^2
\label{dispersion}
\end{equation}
for the dispersion relation of density waves \cite{Bertin2014dyga.book}, where $m$ is the azimuthal wavenumber, $\Omega = \Omega(r)$ is the angular velocity describing the differential rotation of the galactic disk, $v(r) = r \Omega$ and $J(r) = r^2 \Omega$ are respectively the angular rotation velocity and specific angular momentum. $\kappa$, the epicyclic frequency, is given by
\begin{equation}
     \kappa^2 = \frac{2 \Omega}{r} \frac{d}{dr}(r^2 \Omega), 
     \label{epicyclic}
\end{equation}
In \eqref{dispersion}, $\sigma$ represents the sound speed in a fluid disk, while it represents the radial  velocity dispersion in a stellar disk. Although the dispersion formula~\eqref{dispersion} is easiest derived for a gas, the main ingredients, namely that the locally preferred wavenumber $k_1=\pi G \Sigma_B/\sigma^2$ at the neutrally stability point where $Q=\sigma \kappa/\pi G \Sigma_B$ is unity, are still valid for more complex models. 
Stability of the disk to gravitational clumping requires that the Toomre parameter $Q = \frac{\sigma \kappa}{\pi G \Sigma_B} > 1$. A positive $\kappa^2$ implies that the specific angular momentum $r^2 \Omega$ is increasing with $r$, and thus stabilizes the long wavelengths $k \to 0$. This is indeed Rayleigh's criterion for stability of rotating inviscid flows. $k_1(r) = \frac{\pi G \Sigma_B}{\sigma^2}$ is the ``locally" preferred wavenumber, i.e. it is associated with the fastest growing modes.
 
The Toomre parameter $Q$ measures the relative sizes of the destabilizing and the stabilizing effects. We argue that, in the aftermath of an unstable initial state with $Q<1$, there will be a nonlinear feedback in which the gas/stellar disk heats up (the average of the square of the radial velocity fluctuation will increase) until a new nonlinear equilibrium is reached consisting of a pattern with local preferred wavenumber $k_1=\pi G \Sigma_B/\sigma^2$ and with a Toomre parameter of unity. We emphasize that the argument we now present relating the $\Sigma^*$ and $k_0$ in terms of the baryonic mass $M_B$ and the universal acceleration only uses radial averages \cite{Romeo2018Angular} of the local preferred wavenumber and the Toomre parameter. We start from the relations
\begin{align}
\label{eq:k1}                      k_1(r)& =\frac{\pi G \Sigma_B}{\sigma^2}    \\ 
\label{eq:Toomre}                      \pi G \Sigma_B & \lesssim \sigma \kappa=\frac{\sqrt{2}\sigma v_{\infty}}{r} 
\end{align}
since the epicyclic frequency is $\sqrt{2}v_{\infty}/r$ for the flat part of the rotation curve. We want to ``average" these equations with respect to the baryonic density $\Sigma_B$. Multiplying \eqref{eq:k1} by $2\pi r \Sigma_B$ and integrating over $r$ gives 
\begin{equation}
 M_B \bar{k}  = 2 \pi \int k_1(r) \Sigma_B rdr = 2 \pi^2 G \int\frac{\Sigma_B^2}{\sigma^2} r dr,      
    \label{eq:k1integrated}
\end{equation}
where we have used $M_B=\int 2 \pi r \Sigma_B(r)dr$, and defined $\bar{k}$, the average of the local wavenumber $k_1(r)$ by
\begin{equation}
     \bar{k} =\frac{2 \pi}{M_B} \int k_1(r) \Sigma_B rdr.   
     \label{k0eqavg}
\end{equation}
Next, we rearrange, square and then integrate \eqref{eq:Toomre} to obtain
\begin{equation}
     \pi^2 G^2 \int \frac{\Sigma_B^2}{\sigma^2} r dr\lesssim \int_{r_{in}}^{r_{out}} \frac{2 v^2_{\infty}}{r} dr = 2 v_{\infty}^2 \ln\left(\frac{r_{out}}{r_{in}}\right),  
     \label{effectiver1}
\end{equation}
where $r_{out}$ is the ``effective" range of $\Sigma_B$, a distance that contains an $O(1)$ fraction of the baryonic mass and beyond which the surface density is negligible, and $r_{in}$ is the inner scale for the transition from a linear to a flat rotation curve. 

Because $r_{in}$, which we expect to be of the order of $k_0^{-1}$ (the scale for the pattern halo) and $r_{out}$, which we expect to be a few times the baryonic scale length $r_0$, appear in the argument of a logarithm, the following argument is insensitive to their precise values. Combining  \eqref{k0eqavg} and \eqref{effectiver1} we get
\begin{equation}
GM_B \bar{k} \simeq 2 v_\infty^2 \ln\left(\frac{r_{out}}{r_{in}}\right)
\end{equation} 
Multiplying this equation by by Eq.~\eqref{eq:asymptotic} we get 
\begin{equation}
     v_{\infty}^4 \simeq  G M_B \cdot (2\pi G \Sigma^*) \cdot \left(4\left(\frac{\bar{k}}{k_0}\right) \ln\left(\frac{r_{out}}{r_{in}}\right) \right)
     \label{v4M}
\end{equation}
This is the baryonic Tully Fisher relation (BTFR) with the acceleration $a_0=2\pi G \Sigma^*$ times $4\left(\frac{\bar{k}}{k_0}\right) \ln\left(\frac{r_{out}}{r_{in}}\right)$. We will demand that $4\left(\frac{\bar{k}}{k_0}\right) \ln\left(\frac{r_{out}}{r_{in}}\right)$, a quantity that can vary between galaxies, and might depend on their geometry and details of the matter distribution, be approximately equal to 1. Thus the choices that
\begin{equation}
    \Sigma^*=\frac{a_0}{2 \pi G},\quad \bar{k} \sim k_0 \propto \sqrt{\frac{2\pi\Sigma^*}{M_B}}   
    \label{calibration}     
\end{equation}
are entirely reasonable and consistent not only with all known observational data but with the ideas that in the wake of a gravitational instability there is a preferred wavenumber and that the system in an average sense evolves nonlinearly to a state where the Toomre parameter in unity. And of course, these choices are also entirely consistent with MOND whose premise is that the Newtonian gravitational acceleration $GM_B/r^2 \leq a_0$, it should be replaced by its geometric mean with the universal acceleration $a_0$. We also remark that the analog of Eq.~\eqref{v4M} for elliptic galaxies is the Faber-Jackson relation \cite{Faber1976Velocity}. Interestingly, using the small variations of $Q$ across a wide range of galaxies along with {\em independent arguments}, Romeo and coworkers have obtained scaling laws for disk galaxies \cite{Romeo2018Angular,Romeo2020Scaling} that highlight and the role of local gravitational instabilities in galaxy evolution \cite{Romeo2020Scaling,Romeo2020From}.

%%%%%%%%%%%%%%%%%%%%%%%%%%%%%%%%%%%%%%%%%%%%%%%%
%
%		Variational analysis
%
%%%%%%%%%%%%%%%%%%%%%%%%%%%%%%%%%%%%%%%%%%%%%%%%

\section{Galaxies with pattern dark matter} \label{sec:LTG}

We now describe the dynamics of galaxies in the context of our model~\eqref{Lagrangian}. Initially we will work with a prescribed baryonic density distribution $\rho_B$. In subsequent sections, we will build self-consistent models of galaxies by finding appropriate solutions of the collisionless Boltzmann equation  including the effects of pattern dark matter.

Since galaxies are non-relativistic, $v_{\infty} \ll c$, 
the geometry of space-time deviates from the flat Minkowski space at $O(\epsilon)$ where 
$\epsilon=\left(\frac{v_\infty}{c}\right)^2.$ 
We obtain the (Newtonian) limit description through a principled asymptotic expansion in the small parameter $\epsilon$. In a steady state, our system is described by the weak-field metric, 
$
g  = -(c^2+ 2 \phi(\mathbf{x})) dt^2 + (1 - 2\phi(\mathbf{x})/c^2) (dx^2+dy^2+dz^2), 
$ where $\phi(\mathbf{x})$ is the total Newtonian potential. We note that $\psi, \mathbf{x}, \mathbf{k}$
are $O(1)$, the spatial velocity $\mathbf{v} = \frac{d\mathbf{x}}{dt}$ is $O(\sqrt{\epsilon})$, and $\rho_B$, $\Sigma^*$ and $\phi$ are
$O(\epsilon)$.
We can expand the action $\mathcal{S}$ and collect terms in powers of $c$ (equivalently $\epsilon$) to get,  $\mathcal{S} = c^2 \mathcal{S}_1 + \mathcal{S}_2$,
\begin{align}
       \mathcal{S}_1 & = -\int d^3\mathbf{x} dt\,\left[ \Sigma^* k_0^{-3} [(k_0^2-|\nabla \psi|^2)^2 + (\Delta \psi)^2] + \tilde{\rho} V(|\nabla \psi|^2) \right] \nonumber \\
    \mathcal{S}_2 & = \int  d^3\mathbf{x} dt \, \left[\rho_B \left(\frac{\mathbf{v}^2}{2} -\phi\right) -\frac{|\nabla \phi|^2}{8 \pi G} - 2 \phi \Sigma^* k_0^{-3}  \left(|\nabla \psi|^4-k_0^4 \right)  \right. \nonumber \\
    & - \left. 2 \phi\left( \Sigma^* k_0^{-3}  \left(\Delta \psi\right)^2 + \tilde{\rho} V'(|\nabla \psi|^2) |\nabla \psi|^2 \right)\right].
    \label{newtonian-limit}
\end{align}
This formulation is completed by prescribing the potential $V$.

We illustrate the procedure for analyzing the variational equations for the action in~\eqref{Lagrangian} by revisiting the example of spherically symmetric compact clump of matter. {\bf Step 1:} Prescribe $\rho_B(R)$ and solve the variational equations for $\mathcal{S}_1$, i.e. a pattern formation problem. For a compact clump, and a generic potential $V$ with a global minimum at 0,  $\nabla \psi \approx 0$ within the source, so we get the target patterns that were discussed earlier. {\bf Step 2:} With the given $\rho_B$  and $\psi$ computed from the previous step, solve for the gravitational potential $\phi$. For a compact dense clump, $\nabla \psi\approx 0$ where $\rho_B \neq 0$, and outside the clump, $|\nabla \psi| \approx 1, \Delta \psi \approx 2 R^{-1}$. Consequently, we get 
\begin{align}
\Delta \phi & \approx 4 \pi G\left(\rho_B + \frac{8 \Sigma^* k_0^3 R^2}{1 + 2 k_0^2 R^2}\right), \nonumber \\
g_{\text{obs}} & = \nabla \phi \approx \frac{G(M_B+M_P)}{R^2}  
\end{align}
{\bf Step 3:} Solve for the steady state velocity from $\frac{v^2}{R} = g_{\text{obs}}$. 

\subsection{Variational analysis of disk galaxies} \label{sec:eikonal}

To model a disk galaxy, we now carry out these steps in an axisymmetric setting, where all the fields only depend on $r = \sqrt{x^2+y^2}$ and $z$. The matter density $\rho_B(r,z) \approx \Sigma_B(r) \delta(z)$ is concentrated close to the galactic plane $z=0$.
 
 In Step 1, extremizing $\mathcal{S}_1$, we have two contributions, the pattern Lagrangian $\mathcal{S}_P$ which is an integral over all of space, and the interaction Lagrangian $\mathcal{S}_\psi = -2 \pi \int \Sigma_B(r) V(|\psi_r|^2) r dr$ which is an integral over the galactic disk. Off the disk  $\psi$ satisfies the {\em Eikonal equation} $|\nabla \psi| = k_0$, as appropriate for stripe patterns. 
Using  Huygens' principle, we obtain:
 \begin{align}
& \psi(r,z) = \min_{s \geq 0} \left[\psi(s,0) + k_0 \sqrt{(r-s)^2 + z^2}\right]  \nonumber \\
 \Rightarrow \quad & \psi\left[s+  t \cos \theta(s), \pm t \sin \theta(s) \right]  = \psi(s,0) + k_0 t. 
 \label{huygens}
\end{align}
where the second line follows for regions where the  {\em characteristics} $r = s+  t \cos \theta(s), z=\pm t \sin \theta(s))$ do not cross. 

\begin{figure}%  figure placement: here, top, bottom, or page
  \centering
  \includegraphics[width=0.7 \textwidth]{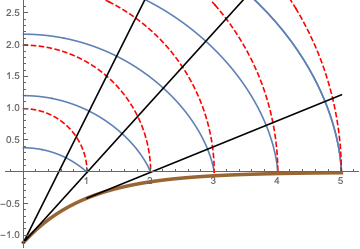} 
  \caption{Huygens' construction -- 
  the phase contours (solid curves in $z > 0$) have a common evolute (solid curve in $z<0$)  and intersect the characteristics (straight lines) orthogonally.  The  contours for $z \leq 0$ are given by reflection.   A spherical target pattern (dashed phase contours) is shown for comparison. The involutes are (approximately) spherical caps with centers off the plane $z=0$.
      }
  \label{construction}
\end{figure}

\begin{figure}%  figure placement: here, top, bottom, or page
  \centering
  \includegraphics[width=0.7 \textwidth]{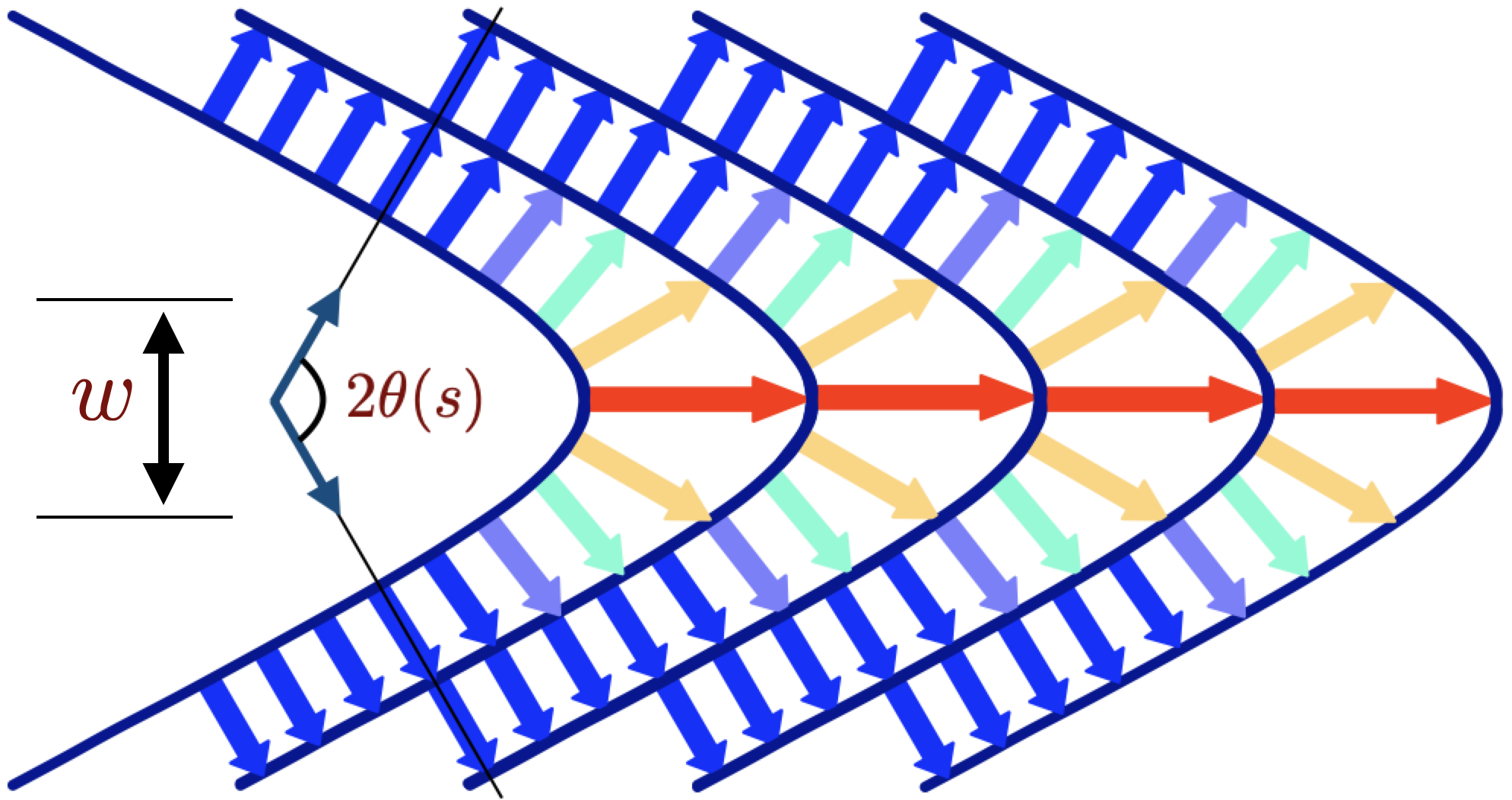} 
  \caption{Phase grain boundary (PGB). There is a jump in $\nabla \psi$ across the PGB. This structure is smooth on the scale $w$, the width of the PGB. The stretching and bending of the phase contours  contribute to an effective surface energy on the grain boundary.}
  \label{fig:PGB}
\end{figure}

The geometry of this construction is illustrated in Fig.~\ref{construction}. The phase fronts  for $z > 0$ (resp. $z < 0$) are the {\em involutes} of a common {\em evolute} $\gamma = (\alpha(s),\mp \beta(s))$ and $k_0^{-1}\psi$ is the local radius of curvature \cite[\S 12]{Rutter2000Geometry}. For the Eikonal solution, $\nabla \psi$ is discontinuous across the galactic plane $z=0$. Indeed, in contrast to the spherical target pattern, the contours given by the involutes intersect the plane $z=0$ at an angle $\theta(s) \neq \frac{\pi}{2}$. This discontinuity in $\nabla \psi$ is regularized as a {\em phase grain boundary} (PGB), a defect well known in patterns, consisting of  a boundary layer across which  $\nabla \psi$ changes smoothly as illustrated in Fig.~\ref{fig:PGB}.  

We can estimate the (surface) energy density of a PGB as follows. Since the boundary layer has width $w$, the curvature and stretch of the phase contours are, respectively, 
$\Delta \psi \sim k_0 w^{-1} \sin \theta(s), k_0^2-|\nabla \psi|^2 \sim k_0^2 \sin^2 \theta(s)$.
Eq.~\eqref{CNred} now implies  
$$
\Sigma_{PGB} \sim w^{-1} k_0^2 \sin^2 \theta(s) + w k_0^4 \sin^4 \theta(s).
$$ 
Optimizing for $w$ gives $w \sim \frac{1}{k_0 \sin \theta(s)}, \Sigma_{PGB} \propto \sin^3 \theta(s)$. A rigorous calculation along these lines yields $\Sigma_{PGB} = \frac{8 \Sigma^*}{3} \sin^3\theta(s)$ \cite{newell1996defects}. Using~\eqref{calibration}, the sum of $\mathcal{S}_\psi$ and the PGB defect energy is 
\begin{equation}
    \label{disk-energy}
    \mathcal{S}_{\text{disk}} = 2 \pi \int \left[\frac{8 \Sigma^*}{3} \sin^3 \theta(s) + \Sigma_B(s) V(k_0^2 \cos^2\theta(s)) \right] s ds.
\end{equation}
We can extremize to get $k_0^2 \Sigma_B(s) V'(\cos^2 \theta(s))=4 \Sigma^* \sin \theta(s)  $, {\em a local relation} between the matter surface density, the characteristic angle $\theta(s)$, and indirectly, also the common evolute $\gamma$. It is important to note that we are equating the grain boundary energy with the effective mass of the pattern dark matter near the galactic plane. 

We can now make an informed choice for the potential $V$. The argument of $V$ is $| \nabla \psi|^2 =k_0^2 \cos^2 \theta(s) \leq k_0^2$ within the galactic disk. To ensure this is always true, i,e. that  $|\nabla \psi|^2 \leq k_0^2$ in the presence of matter, a natural choice is the "infnite-well" potential 
$$
V(|\nabla\psi|^2) = \begin{cases} 0 & |\nabla \psi|^2 \leq k_0^2 \\ + \infty & \mbox{ otherwise } \end{cases}
$$
This potential is clearly not smooth, or even strictly convex. In convex optimization, a canonical ``replacement" of the infinite well potential without these shortcomings, and one which is amenable to numerical minimization, is the {\em log barrier function} 
\begin{equation}
V = -V_0 \ln(k_0^2-|\nabla \psi|^2)
\label{log-barrier}
\end{equation}
which leads to interior point methods in convex optimization \cite{boyd2004convex}.

We will use the log barrier function to model the interaction and take $V_0$ to be an $O(1)$ constant. Putting everything together, we have the leading order (in $\epsilon = (v_\infty/c)^2$ small and $k_0 \sqrt{r^2+z^2}$ large) solution of the variational equations for~\eqref{Lagrangian}:
\begin{align}
(r,z) & = (s + t \cos \theta(s), \pm t \sin \theta(s)), \nonumber \\
\gamma & = \left(s + \frac{\cos \theta(s) \sin \theta(s)}{\theta'(s)}, \frac{\sin^2 \theta(s)}{\theta'(s)}\right) \nonumber \\
    \Sigma_B(s) & = \frac{4 \Sigma^*}{V_0} \sin^3 \theta(s) \nonumber \\
    |\nabla \psi| & \simeq k_0, \quad \Delta \psi \simeq 2 k_0 \left(t - \sin \theta(s) /\theta'(s)\right)^{-1} \approx 2k_0/\sqrt{r^2+z^2}, \nonumber \\
    \Delta \phi & = \Delta (\phi_{B} + \phi_P) \simeq  4 \pi G \left[\Sigma_B(r) \delta(z) + 2 \Sigma^* k_0^{-3} (\Delta \psi)^2\right], \nonumber \\
    v^2 & =  r  \partial_r \phi(r,0) =r  \partial_r \phi_B(r,0) + r \partial_r \phi_P(r,0).
    \label{galaxy}
\end{align}

The leading order solution of $\psi$ is given by~\eqref{huygens} as long as the curvature $\Delta \psi \lesssim k_0^2$, the curvature scale of the PGB. This is consistent with a `cored dark halo' since the $\rho_P \sim |\Delta \psi|^2$ is bounded and not divergent as in the cuspy NFW profile. 

In the last line in~\eqref{galaxy}, we have included $\nabla \phi$, the gradient of the Newtonian potential, which comes from $\mathcal{S}_2$ while ignoring the ``fifth force" that arises from the gradient of the interaction term $\tilde{\rho} V(|\nabla \psi|^2)$ which is formally of higher order. This is justified by a separation of scales. The effect of displacements of a star on the interaction term, which is ``large scale", is suppressed by the smallness of ratio of the stellar radius to $k_0^{-1}$, the scale on which $|\nabla \psi|$ varies. On the other hand, this term can be important in situations where two distinct clumps of matter, each on the scale of $k_0^{-1}$ are interacting, for instance, between a galaxy and its satellites or between galaxies in a cluster.

%%%%%%%%%%%%%%%%%%%%%%%%%%%%%%%%%%%%%%%%%%%%%%%%
%
%		Rotation curves
%
%%%%%%%%%%%%%%%%%%%%%%%%%%%%%%%%%%%%%%%%%%%%%%%%

\subsection{The rotation curves for Kuzmin and Exponential disks} \label{sec:RC_disks}

We will first consider dynamically cold, i.e. purely rotation supported disk galaxies with no significant random motions or 3d structure, i.e. no bulge. In this case $v(r)$ is the azimuthal rotation velocity for circular orbits in the effective (baryonic + pattern) gravitational field of the galaxy. For such galaxies,~\eqref{galaxy} implies the  Freeman limit $\Sigma_B \leq \frac{4 \Sigma^*}{V_0}$ \cite{McGaugh1995galaxy}. We will discuss pressure supported systems in subsequent sections. 

The second equation in~\eqref{galaxy} expresses the common evolute $\gamma$ in terms of $\theta(s)$ which in turn is given by $\Sigma_B$. This connects the {\em local} matter distribution $\Sigma_B$ and the pattern `halo' \cite{Sancisi2004visible}. 
We can also prescribe $\gamma$ and use it to compute $\Sigma_B, \psi,\phi$ and $v$. A natural {\em critical} case is when the evolute degenerates to a single point $(0,-z_0)$, so that $\theta(s) = \arctan(\frac{z_0}{s})$ and $\Sigma_B(s) = \frac{4\Sigma^*}{V_0}(1 + s^2/z_0^2)^{-3/2}$, corresponding to a Kuzmin disk. 
It is remarkable that the surface density of a Kuzmin disk, a natural model for galactic disks, arises from the surface energy $\propto \sin^3\theta(s)$ relation for PGB defects, a formula that was originally derived in a totally different context of patterns  \cite{newell1996defects}. 

The mass of this `critical' Kuzmin disk, $M_B = 8 \pi \Sigma^* z_0^2/V_0$, is determined by $z_0$, the length-scale in the evolute. The phase is given by $\psi(r,z) = k_0(r^2 + (|z| + z_0)^2)^{1/2}$ and the curvature of the contours is $1/(r^2 + (|z| + z_0)^2)^{1/2} \leq z_0^{-1}$ so the eikonal approximation for the phase is valid for all $(r,z)$. The Newtonian potential of the Kuzmin disk is
$$
\phi_{\text{bar}}(r,z) = -\frac{G M_B}{\sqrt{r^2 + (z_0 + |z|)^2}} = - \frac{4}{V_0} \frac{(2 \pi G \Sigma^*) z_0^2}{\sqrt{r^2 + (z_0 + |z|)^2}}.
$$
The halo contribution to the potential is given by solving $-\Delta \phi_{\text{halo}} \approx 8 \pi G \Sigma^* k_0^{-1} (r^2 + (z_0 + |z|)^2)^{-1}$. We can solve for the potential, on the plane $z=0$, using the Fourier-Bessel transform \cite{BT08}, to get
$$
\phi_{\text{halo}}(r,0) \approx   V_0^{-1/2} \left(\ln(r^2 + z_0^2) + 2 K\left(-\frac{r^2}{z_0^2}\right)\right),
$$
where $K$ is the complete elliptic integral of the first kind~\cite{Abram_Stegun}. The leading order expression for potential $\phi(r,0)$ and for the azimuthal rotation velocity can now be computed to yield:
\begin{align}
    \label{kuzmin-pot}
    \frac{\phi(r,0)}{2 \pi G \Sigma^* z_0} & \approx -\frac{4}{V_0\sqrt{1+\xi^2}}  +  V_0^{-1/2} \ln(1+\xi^2) + \cdots, \nonumber \\ 
    \frac{v^2(r)}{2 \pi G \Sigma^* z_0}  & \approx \frac{4 \xi^2}{V_0(1+\xi^2)^{3/2}} + 2 V_0^{-1/2}\frac{\xi^2}{{1+\xi^2}} + \cdots,  \nonumber \\
    \frac{g_{\text{obs}}}{2 \pi G \Sigma^*} & \approx \frac{4 \xi}{V_0(1+\xi^2)^{3/2}} + 2 V_0^{-1/2}\frac{\xi}{{1+\xi^2}} + \cdots 
\end{align}
where $\xi = r/z_0$ is the scaled radius, and the initial terms are the (non-dimensional) baryonic contributions to the potential ($\phi_{\text{bar}}$), velocity ($v_{\text{disk}}$) and acceleration ($g_{\text{bar}}$).  The asymptotic velocity $v_\infty^2 = 4 \pi V_0^{-1/2} G \Sigma^* z_0 \equiv (G M_B a_0)^{1/2}$ where $a_0 = 2 \pi G \Sigma^*$. Independent of the scale $z_0$, the critical Kuzmin disks in our theory satisfy a radial acceleration relation (RAR) since both $\frac{g_{\text{bar}}}{a_0}$ and $\frac{g_{\text{obs}}}{a_0}$ only depend on the combination $\xi = r/z_0$.  We will return to this point in Sec.~\ref{sec:RAR}.

While the Kuzmin disk is a useful model, most real galaxies are exponential disks \cite{Freeman1970disks}. Interestingly, exponential disks also arise naturally in our theory. From~\eqref{galaxy}, a ``limiting" case for a cored halo corresponds to  $\theta'(s) = z_0^{-1} \sin \theta(s)$ which ensures that  $\Delta \psi \geq 2 k_0 z_0^{-1}$. Solving for $\theta(s)$ and computing the corresponding density $\Sigma_B$ using~\eqref{galaxy}, we get,
\begin{equation}
\Sigma_B(s) = \frac{4 \Sigma^* A^3}{V_0} \frac{e^{-3 s/z_0}}{(1+A^2 e^{-2 s/z_0}/4)^3}.
\label{exp-disk}
\end{equation}
For $A \lesssim 1$, this is the baryonic density of an exponential disk $\Sigma_B = \Sigma_0 e^{-s/r_0}$ with $\Sigma_0 = \frac{4 \Sigma^* A^3}{V_0}, r_0 = \frac{1}{3}z_0$, suggesting that the self-organizing processes underlying our model might naturally produce exponential disks if the dynamics drive the phase curvatures to a constant (maximal) value on the galactic plane.

\begin{figure} \sidecaption %  figure placement: here, top, bottom, or page
  \centering
  \includegraphics[width=0.7 \textwidth]{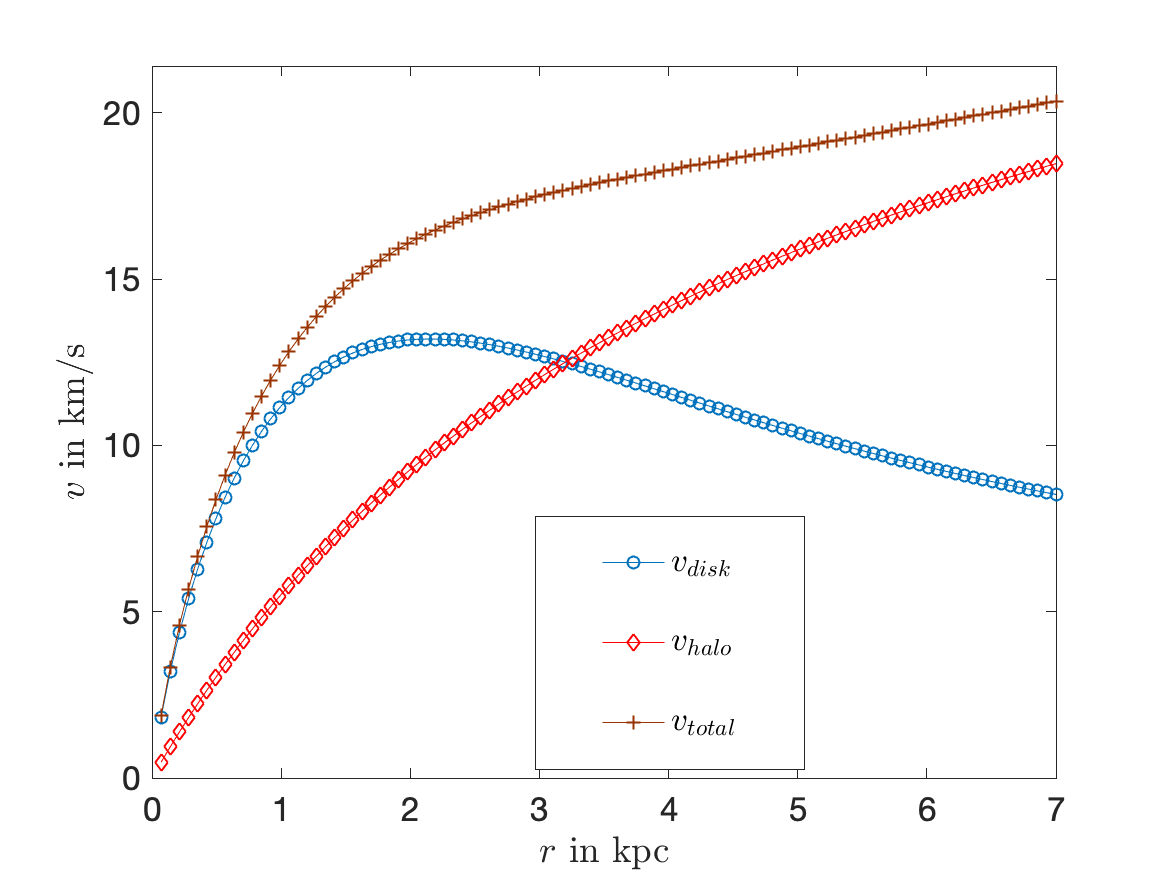} 
  \caption{   Computed rotation curves. $a_0 = 3600 \,\text{km}^2 \text{s}^{-2} \text{kpc}^{-1}$. We approximate the 
    pattern DM density $2 \Sigma^* k_0^{-1} (\Delta \psi)^2$ by the $\ell = 0$ mode. For these parameters $v_{\infty} \approx 35$ km/s so the rotation curve continues to rise over many scale lengths. Compare Fig.~7 in Ref.~\protect{\cite{DiPaolo2019universal} }}
  \label{rot-curves}
\end{figure}

Our theory can calculate the rotation curves for any prescribed (LSB) surface density $\Sigma_B(s) \leq 4 \Sigma^*/V_0$ including exponential disks. We henceforth set $V_0 = 4$. Combining~\eqref{exp-disk} and~\eqref{calibration} we obtain $k_0 z_0 =  48 \,A^{-3/2}$.  Fig.~\ref{rot-curves} shows the numerically obtained rotation curves for a model exponential disk with $r_0 = 1 \,\text{kpc}, M_B = 10^8 M_\odot$ corresponding to $A \approx 1/2$.
The rotation curve computed from our theory shows that rises slowly, and continues to rise beyond $7 r_0$. The shape of the curve as well as the scale of the velocity is in good qualitative agreement with the observational curves in \cite{DiPaolo2019universal}.

%%%%%%%%%%%%%%%%%%%%%%%%%%%%%%%%%%%%%%%%%%%%%%%%
%
%		Elliptical galaxies and pressure support
%
%%%%%%%%%%%%%%%%%%%%%%%%%%%%%%%%%%%%%%%%%%%%%%%%

\section{Self-organization and dynamical equilibria}  \label{sec:spherical}

 In the previous section we analyzed the rotation curves of {\em cold disk galaxies  with prescribed surface density profiles} $\Sigma_B(s)$ with a $k_0$ that is determined by demanding that the differentially rotating disk be marginally stable. In this section, we will remove these assumptions and start building self-consistent solutions that better reflect the dynamical processes governing galaxies.
 
We being in section~\ref{sec:dynamical} by discussing a dynamical mechanism, independent of the stability of differentially rotating disks, that determines $k_0$ through a nonlinear eigenvalue problem. A byproduct of this analysis if the emergence of a non-dimensional parameter $\eta$ that distinguishes high surface brightness (HSB) galaxies from low surface brightness (LSB) galaxies in terms of the dependence of $k_0$, and hence the nature of the pattern halo, on the baryonic mass $M_B$  of the galaxy. We continue in section~\ref{sec:DF} by building self-consistent spherical galaxies through solutions for the appropriate galaxy distributions functions \cite{BT08}. The solutions are dynamically self-organized states characterized by two parameters, the total baryonic mass $M_B$ and an exponent $\gamma$ governing the density falloff as a function of the distance from the center of the galaxy. The analysis leads to the Faber-Jackson relation \cite{Faber1976Velocity}, and to two distinct types of solutions, bright/compact galaxies with $\gamma \geq 4$ and dim/diffuse galaxies with $\gamma \approx 3$, that are distinguished by the (projected) central surface density. We then discuss the fundamental plane for our model spherical galaxies in section~\ref{sec:FP}, and show that there are distinct relations for luminous and for dim galaxies, in agreement with observations \cite{Gudehus1991Systematic}.  
 
\subsection{A dynamical mechanism for selecting $k_0$} \label{sec:dynamical}

A central consequence of our theory is that the phase $\psi$ is ``slowly-varying", i.e. $O(1)$ changes of the phase occur on a scale $k_0^{-1}$ which is much larger than the sizes of the stars and other ``condensed" objects. Consequently, we can replace $\rho_B$ by a smoothed version $\tilde{\rho} = K \star \rho_B$, obtained by convolving with a (normalized) Gaussian kernel $K$ with width $k_0^{-1}$, in the Largangian~\eqref{Lagrangian}.  The terms in the Lagrangian involving the phase $\psi$ are $\mathcal{S}_P$ and $\mathcal{S}_\psi$, so that $\psi$ is determined by extremizing 
$$
\mathcal{S}_P + \mathcal{S}_\psi \simeq   \int \left(\frac{\Sigma^* c^2}{k_0^3} \left\{(k_0^2-\nabla^\mu \psi \nabla_\mu \psi)^2 + (\nabla^\mu \nabla_\mu \psi )^2\right\} + \tilde{\rho} c^2 V\left[k_0^{-2} \nabla^\mu \psi \nabla_\mu \psi \right] \right)\sqrt{-g} \ d^4x. 
$$
While this averaging $\rho \to \tilde{\rho} = K \star \rho$ does not affect the action, it does give a principled approach to relating $k_0$ to the baryonic density $\rho_B$, as we now argue.

A self-dual reduction for the energy functional $\mathcal{S}_P + \mathcal{S}_\psi$ is given by setting a ``dominant balance" through matching the various terms in the energy. Accounting for the signs of the curvature $\Delta \psi$, the ``stretching" $k_0^2 - |\nabla \psi|^2$ and the (smoothed) density $\tilde{\rho}$, we posit
$$
\nabla^\mu \nabla_\mu \psi = \sqrt{\frac{\tilde{\rho} k_0^3}{\Sigma^*} V\left[\frac{ \nabla^\mu \psi \nabla_\mu \psi}{k_0^2}\right]} - (k_0^2-\nabla^\mu \psi \nabla_\mu \psi)
$$
where we are taking the positive square root. This equation has terms with similar spatial variations since $\tilde{\rho}$ is smooth on the scale $k_0^{-1}$. Conversely, this equation will have an ``unbalanced" rapidly varying term if we use the true baryonic density $\rho_B$ instead of its smoothed average $\tilde{\rho}$. An alternative viewpoint is that the pattern field $\psi$ is ``universal" in that it only depends on the coarse-grained (and hence large-scale/nonlocal) features of the density distribution, and not on the local/microscopic details of $\rho_B$.

The Hopf-Cole transformation $\psi = -\ln \Psi$  yields
$$
- \frac{\nabla^\mu \nabla_\mu \Psi}{\Psi} + \frac{\nabla^\mu \Psi \nabla_\mu \Psi}{\Psi^2} =  \sqrt{\frac{\tilde{\rho} k_0^3}{\Sigma^*} V\left[\frac{ \nabla^\mu \Psi \nabla_\mu \Psi}{k_0^2 \Psi^2}\right]} +  \frac{\nabla^\mu \Psi \nabla_\mu \Psi}{\Psi^2}-k_0^2.
$$
Rearranging gives the (nonlinear) Schr\"odinger equation
$$
(-\nabla^\mu \nabla_\mu + W)\Psi = -k_0^2 \Psi
$$
where 
$$
W = -  \sqrt{\frac{\tilde{\rho} k_0^3}{\Sigma^*} V\left[\frac{ \nabla^\mu \Psi \nabla_\mu \Psi}{k_0^2 \Psi^2}\right]}
$$
is the (attractive) potential. $-k_0^2$ is then the ground state energy, and the phase $\psi$ is determined by the negative logarithm of the ground state wavefunction. Provided that the potential $W$ supports a bound state, this procedure is well defined since the ground state wavefunction is nowhere vanishing and real (WLOG), so the logarithm is thus well defined. The potential $W$ depends on $k_0$ and also the Hopf-Cole transform of the phase $\Psi = e^{-\psi}$, so this is a self-consistent determination for $k_0$. 

We illustrate this approach for a point mass $M_B$. To determine the scaling of $k_0$, we can neglect the details of the dependence of $W$ on $V$, an $O(1)$ quantity that varies ``slowly", i.e. on the scale $k_0^{-1}$. These details can affect the numerical prefactors, but not the scaling dependence of $k_0$ on $M_B$. After smoothing, we have $\tilde{\rho} \sim M_B k_0^3$ so the Schr\"odinger problem (approximately) corresponds to a particle in a 3d spherical box with radius $k_0^{-1}$ and depth $\sqrt{\frac{M_B}{\Sigma^*}} k_0^3$. Rescaling with $\xi = k_0 R$ gives the eigenvalue problem
\begin{align}
-\Psi''(\xi) - \frac{2}{\xi} \Psi'(\xi) + \Psi(\xi)& = \begin{cases} \sqrt{\frac{M_B}{\Sigma^*}} k_0 \Psi(\xi) & 0 \leq \xi \leq 1 \\
0 & \xi > 1 \end{cases} \nonumber \\
\Psi'(0) = 0,  \quad \Psi(\xi) & \to 0 \mbox{ as } \xi \to \infty
\end{align}
A straightforward calculation now gives $k_0 \approx 2.2 \sqrt{\Sigma^*/M_B}$. More generally, independent of the details of the smoothing, and of the potential $V$, a similar rescaling argument will result in an nonlinear eigenvalue problem with a single parameter $\sqrt{\frac{M_B}{\Sigma^*}} k_0$ from dimensional considerations. We therefore, generically, will get 
$
k_0 =c \sqrt{\Sigma^*/M_B}
$
for an $O(1)$ constant $c$. This argument is directly inspired by the mechanisms that dynamically determine the wavelength of target patterns in the Belouzov-Zhabotinsky reaction \cite{morris1996spatio}, wherein the pattern wavenumber is determined by the ground state energy of an appropriate Schr\"odinger operator \cite{Kopell1981Target}. In our context, this argument suggests a dynamical mechanism for the origin of $k_0$ and justifies the choices motivated by stability considerations in Sec.~\ref{sec:stability}.

If the mass distribution itself has a length scale $r_0$, then we can no longer assert that $
k_0 =c \sqrt{\Sigma^*/M_B}
$
from dimensional analysis. In this case, we get $\tilde{\rho} \sim M_B a^{-3}$ where $a = \max(r_0,k_0^{-1})$, and the corresponding Schr\"odinger problem is given by a box potential of radius $a$ and depth $W_0 =\sqrt{\frac{M_B}{\Sigma^*}} k_0^{3/2} a^{-3/2}$. A straightforward calculation shows that a box potential in 3d needs to be sufficiently deep, $W_0 a^2 \geq \frac{\pi^2}{4}$, in order to support a bound state. The ground state energy is given by 
\begin{equation}
a^2 k_0^2 \approx \left(\sqrt{\frac{M_B a k_0^3}{\Sigma^*}} - \frac{\pi^2}{4}\right)^2 \quad \mbox{ if } \sqrt{\frac{M_B a k_0^3}{\Sigma^*}} - \frac{\pi^2}{4} > 0.
\label{bound-state}
\end{equation}
More generally, as shown in~\eqref{ball-k0} in Appendix B, for a 3d potential with depth $V_0$ and length scale $a$ there is a critical value $\eta^*$, dependent on the details of the potential, such that $a^2k_0^2 \propto (V_0 a^2 - \eta^*)^2$.

To analyze the condition in~\eqref{bound-state}, we define the (non-negative) dimensionless quantities $\zeta = r_0k_0$ and $\eta = \sqrt{\frac{M_B}{\Sigma^* r_0^2}}$ so that $ak_0  = \max(1,r_0 k_0) = \max(1,\zeta)$ and $\eta = \eta \zeta \max(1,\zeta^{1/2}) - \frac{\pi^2}{4}$. In terms of $\eta$ and $\zeta$ we get the condition
$$
\frac{\pi^2}{4} + \max(1,\zeta) = \eta \zeta \max(1,\zeta^{1/2}),
$$
where $\eta$ is given and we are trying to solve for $\zeta$.

\begin{figure}[htbp] %  figure placement: here, top, bottom, or page
  \centering
  \includegraphics[width=0.8 \textwidth,trim={0cm, 1cm, 0cm, 2cm}, clip]{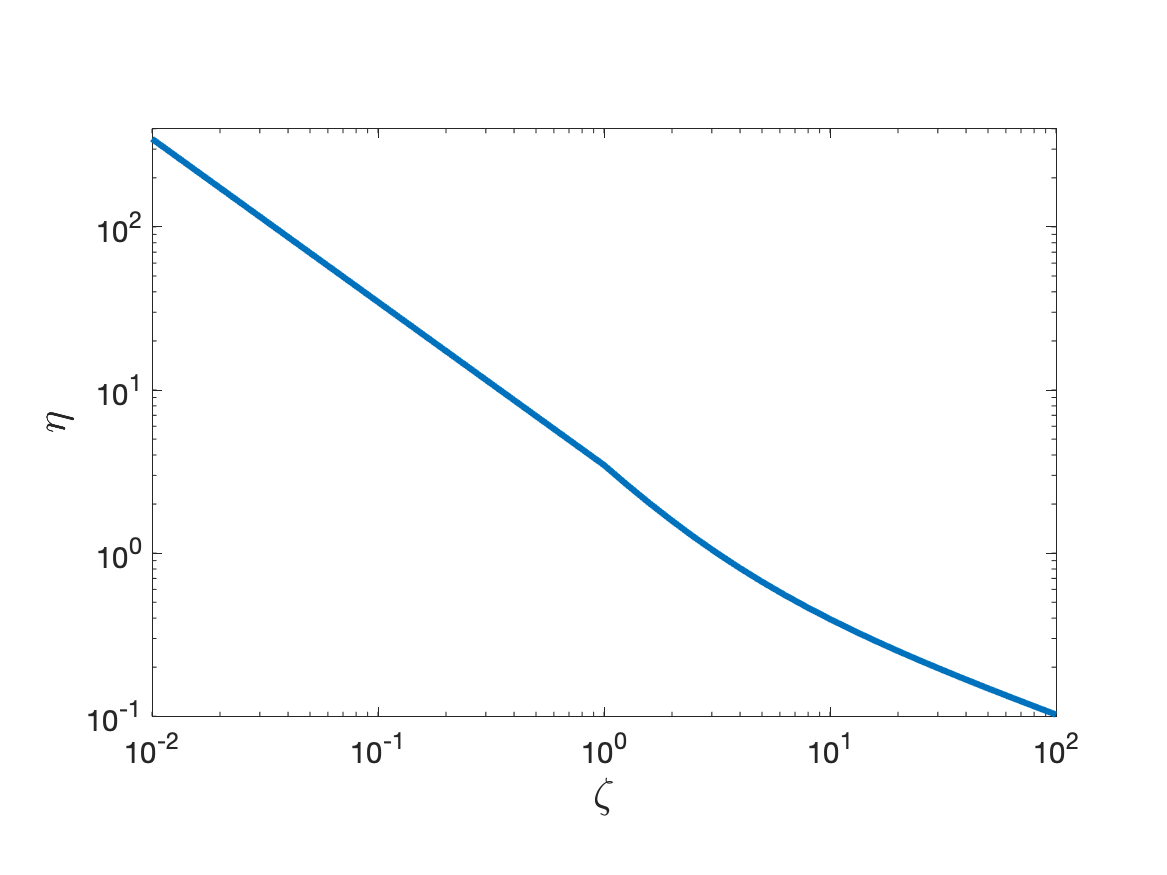} 
  \caption{A plot of $\eta$ as a function of $\zeta$. The curve is strictly monotone with behavior $\eta \sim \zeta^{-1}$ for small $\zeta$ and $\eta \sim \zeta^{-1/2}$ for large $\zeta$.}
  \label{fig:conditions}
\end{figure}

 Fig.~\ref{fig:conditions} is a plot of the ratio $\displaystyle{\frac{\pi^2/4 + \max(1,\zeta)}{\max(\zeta,\zeta^{3/2})}}$ as a function of $\zeta$. This ratio is a monotonic function of $\zeta$ with a range $(0,\infty)$, showing the existence of a unique $\zeta$ that satisfying this condition for any given $\eta$. $\zeta$ is a continuous function of $\eta$, but the nature of the solution changes at the ``break" $\zeta=1$, i.e. depending on whether or not $\eta \geq \eta_c = \frac{\pi^2}{4} + 1$.

Although the precise details, for instance the value of $\eta_c$ will be different for different potentials, the overall conclusions will hold generically. As we will argue below, $\eta > \eta_c$ corresponds to Luminous/High surface brightness (HSB) galaxies, that have a solution  for $\zeta = k_0 r_0$ in the range $\zeta < 1$ that is given by $\zeta = \eta_c/\eta$. In particular, the halo radius $k_0^{-1} \sim \frac{\eta}{\eta_c} r_0 \gg r_0$. Observationally, this would be interpreted as an inner ``Newtonian" region $R \lesssim k_0^{-1}$ where the dynamically inferred mass is dominated by the baryonic density $\rho_B$ and an outer ``MOND" region where the dynamically inferred mass is dominated by the contribution of the pattern field $\psi$. The boundary between these regions is determined by an acceleration scale $GM_B k_0^2 \simeq a_0$ in agreement with the MOND phenomenology. 

Conversely, for a dim/Low surface brightness (LSB) galaxy, $\eta < \eta_c$, the solution is in the range $\zeta >1$ and given by the unique solution of $(\eta_c -1) + \zeta = \eta \zeta^{3/2}$. As $\eta \to 0$, $\zeta \approx \eta^{-2}$ so that $k_0^{-1} \ll r_0$. In this case, the effects of the pattern halo will be evident down to the scale $k_0^{-1}$, again in agreement with the MOND phenomenology. 

Taken together, these results imply
\begin{equation}
k_0 \sim \sqrt{\frac{\Sigma^*}{M_B}} \max\left(\eta_c, \sqrt{\frac{\Sigma^*r_0^2}{M_B}}\right)
\label{mass_k0}
\end{equation}
Note that we have obtained this relation with the implicit assumption that $\tilde{\rho}$ is (roughly) isotropic, and it is not immediately clear that these relations should also hold for mass distributions, like a razor-thin disk, where the aspect ratio between the length scales in the different directions can be substantially different from one. We consider this issue in Appendix B.

%%%%%%%%%%%%%%%%%%%%%%%%%%%%%%%%%%%%%%%%%%%%%%%%
%
%		Galaxy DF
%
%%%%%%%%%%%%%%%%%%%%%%%%%%%%%%%%%%%%%%%%%%%%%%%%

\subsection{Galaxy distribution functions with pattern DM} \label{sec:DF}

Since galaxies contain a large number of individual starts ($N \approx 10^{11}$ for the Milky Way), an adequate description of the collective dynamics of the stars in a galaxy is given by the galaxy distribution function $f(\mathbf{x},\mathbf{v},t)$ describing the phase space (baryonic) density near location $\mathbf{x}$ and velocity $\mathbf{v}$ \cite{BT08}.  The evolution of the distribution function (DF) is given by the collisionless Boltzmann equation (CBE)
\begin{align}
\frac{\partial f}{\partial t} & = - \mathbf{v}\cdot \nabla_{\mathbf{x}} f + \nabla_{\mathbf{x}}\phi \cdot \nabla_{\mathbf{v}} f,  \nonumber \\
\rho_B(\mathbf{x},t) & = \int f(\mathbf{x},\mathbf{v},t) \, d^3 \mathbf{v}, \nonumber \\
\Delta_{\mathbf{x}} \phi & = 4 \pi G (\rho_B + \rho_P),
\label{self-consistent}
\end{align}
where, in a spherically symmetric setting, $\rho_P$ is determined by~\eqref{halo-density}.

By the Jeans theorem any function $f(\mathbf{x},\mathbf{v}) = f(I_1(\mathbf{x},\mathbf{v}),I_2(\mathbf{x},\mathbf{v}),\ldots)$, where $I_1,I_2,\ldots$ are integrals of motion, yields a stationary solution collisionless Boltzmann equation, i.e. describes a state of dynamical equilibrium for the galaxy \cite{BT08}. The (specific) energy $\mathcal{E} = \phi + \frac{|\mathbf{v}|^2}{2}$ is an integral of motion, and for any non-negative function $g(\mathcal{E}) \geq 0$, the {\em ergodic distribution function} $f(\mathbf{x},\mathbf{v}) = g\left(\phi(\mathbf{x}) + \frac{|\mathbf{v}|^2}{2}\right)$ gives a solution of the CBE. A simple and well-motivated DF is the {\em isothermal sphere}, given by 
$$
f = \frac{\rho_0}{(2 \pi \sigma^2)^{3/2}}\exp\left(-\frac{\mathcal{E}}{\sigma^2}\right) \quad \Rightarrow \quad \rho(\mathbf{x}) = \rho_0 \exp\left(-\frac{\phi(\mathbf{x})}{\sigma^2}\right).
$$

In pure Newtonian gravity ($\rho_P =0$), a self-consistent solution for an isothermal DF is the {\em singular isothermal sphere} given by
\begin{equation*}
\phi(\mathbf{x}) = 2 \sigma^2 \ln(|\mathbf{x}|), \quad \rho_0 = \frac{\sigma^2}{2 \pi G}
\end{equation*} 
This gives a baryonic density $\rho_B(R) = \frac{\sigma^2}{2 \pi G R^2}$ that is not normalizable and has infinite mass.

In the context of our model, with the inclusion of pattern DM, an isothermal DF can have finite mass and therefore describe real galaxies, as we now show. Self-consistent spherical solutions are described by two parameters, the total mass $M_B$ and the velocity dispersion $\sigma^2$. Our first goal in this section is to construct galaxy distribution functions that describe self-consistent spherical galaxies also incorporating pattern DM. 

We will also discuss potential observational tests for the pattern DM hypothesis. The (central) velocity dispersion can be deduced from the Doppler broadening of spectral lines. From astronomical observations, we can also measure the total luminosity $L$, the equivalent radius $r_e$, i.e. the radius of the region in the sky that emits half of the total light, and the effective surface brightness $I_e$, the average intensity of the region within the equivalent radius. The various quantities are not independent, and observations show a tight scaling relation between $I_e,r_e$ and $\sigma$, called the {\em fundamental plane} (the relation between $\ln I_e, \ln r_e$ and $\ln \sigma$ is linear). Our second goal in this section is to discuss the (analog of the) fundamental plane in our theory.

For $R \lesssim k_0^{-1}$, the contribution of the pattern density $\rho_P$ is small, and consequently, the baryonic density should correspond to the singular isothermal sphere, so that, 
$$
M_B(R) \simeq 4 \pi \frac{\sigma^2}{2 \pi G} \int_0^{R} \frac{1}{\zeta^2} \zeta^2 d\zeta = \frac{2\sigma^2 R}{G}.
$$ 
The corresponding Newtonian potential is 
$$
\phi_B(R) = \int^R \frac{GM_B(\zeta)}{\zeta^2} d\zeta  = 2 \sigma^2 \ln(k_0 R) + \phi_0
$$
where $\phi_0$ is an arbitrary constant of integration equal to the choice of the potential $\phi$ at $R=k_0^{-1}$. The arbitrary constant $\phi_0$ determines the normalization $\rho_0$ in the relation $\rho_B = \rho_0 e^{-\phi/\sigma^2}$ and choosing $\phi_0=0$ gives $\rho_B = \frac{\sigma^2 k_0^2}{2 \pi G}  e^{-\phi/\sigma^2}$ for all $R$. 
For $R \gtrsim k_0^{-1}$, $\rho_P \gtrsim \rho_B$ and we get 
\begin{align*}
\rho_P & \approx \frac{4 \Sigma^*}{k_0 R^2} \nonumber \\
\phi(R) & \approx \frac{16 \pi G \Sigma^*}{k_0} \ln(k_0 R) + \int_{k_0^{-1}}^R \frac{G M_B(\zeta)}{\zeta^2} d\zeta \nonumber \\
M_B(R) & =   \frac{2\sigma^2}{k_0G} +  \frac{2 \sigma^2 k_0^2}{G}  \int_{k_0^{-1}}^R \exp\left(-\frac{\phi(\zeta)}{\sigma^2}\right) \zeta^2 d\zeta
\end{align*}
An approximate self-consistent solution for the baryonic density is therefore given by 
\begin{equation}
\rho_B(R) = \frac{\sigma^2}{2 \pi G r_0^2} \begin{cases} (r_0/R)^2 & R \ll r_0 \\  (r_0/R)^\gamma & R \gg r_0  \end{cases} 
\label{elliptic_density}
\end{equation}
where $\gamma = \frac{16 \pi G \Sigma^*}{k_0 \sigma^2}$ is a dimensionless parameter. Here $r_0$ is a length scale associated with the baryonic distribution, corresponding to the ``break" between the $R^{-2}$ distribution for small $R$ and the $R^{-\gamma}$ distribtuion for large $R$. We can nondimensionalize and rearrange to get
\begin{equation}
\gamma k_0 r_0 = \frac{16 \pi G \Sigma^* r_0}{\sigma^2}
\label{scaling_regime}
\end{equation}
Smoothly interpolating between the limiting behaviors in~\eqref{elliptic_density}, we will set
\begin{equation}
\rho_B(R) \approx  \frac{\sigma^2}{2 \pi G R^2 (1+R/r_0)^{\gamma-2}}, 
\label{interpolation}
\end{equation}
We emphasize that Eq.~\eqref{interpolation} is an approximation, and a more accurate determination of the baryonic density $\rho_B$ follows from solving the system~\eqref{self-consistent} with $\rho_P$ given by~\eqref{halo-density}. We prefer to work with the approximation in~\eqref{interpolation} since (i) it is correct, including the numerical prefactors, in the limits $R \ll r_0$ and $R \gg r_0$, and (ii) it allows for the following analytic approach to characterizing the self-consistent solutions of~\eqref{self-consistent}. 

 For the total baryonic mass to be finite, we need $\gamma > 3$, and this automatically guarantees the consistency requirement that $\rho_B \ll \rho_P$ for $R \gg k_0^{-1}$. The mass within a (3d) sphere of radius $R$ is given by straightforward integration
\begin{equation}
M_B(R) = \frac{2 \sigma^2}{G} \int_0^R \frac{d\zeta} {(1+ \zeta/r_0)^{\gamma-2}} =  \frac{2 \sigma^2r_0}{G (\gamma-3)}\left[1-\frac{1}{(1+R/r_0)^{\gamma-3}}\right] 
\label{massltR}
\end{equation}
for $\gamma > 3$. %Integrating~\eqref{elliptic_density} over all space we get 
Consequently, the total baryonic mass of the galaxy is given by
\begin{equation}
M_B = \frac{2 \sigma^2 r_0}{G (\gamma-3)}. %= \frac{\gamma}{(\gamma-3)}\frac{\sigma^4}{4 G a_0},
\label{mass}
\end{equation}
Combining Eqs.~\eqref{mass}, \eqref{scaling_regime} and $k_0 = \eta_c \sqrt{\Sigma^*/M_B}$ from \eqref{mass_k0}, we have
\begin{align}
r_0 & = \frac{\eta_c \gamma(\gamma-3)}{32 \pi} \sqrt{\frac{M_B}{\Sigma^*}} = \frac{\eta_c^2 \gamma (\gamma-3)}{32 \pi} \cdot \frac{1}{k_0} \nonumber \\
\sigma^4 & = 2 \pi \left(\frac{8}{\eta_c \gamma} \right)^2  G M_B a_0 
\label{spherical-soln}
\end{align}
where we have used~\eqref{scaling_regime} and $a_0 = 2 \pi G \Sigma^*$ as defined earlier. We have thus recovered the Faber-Jackson relation \cite{Faber1976Velocity} between the baryonic mass and the 4th power of the dispersion. The scatter in this relation comes from the (potential) variation in $\gamma$. 

If we define  $R_{1/2}$ as the radius of the sphere that contains half of the total baryonic mass, it follows from~\eqref{massltR} that 
\begin{equation}
\frac{R_{1/2}}{r_0} = 2^{\frac{1}{\gamma-3}} -1 
\label{scale-length}
\end{equation}

\begin{figure}[htbp] %  figure placement: here, top, bottom, or page
   \centering
   \includegraphics[width=0.7 \textwidth]{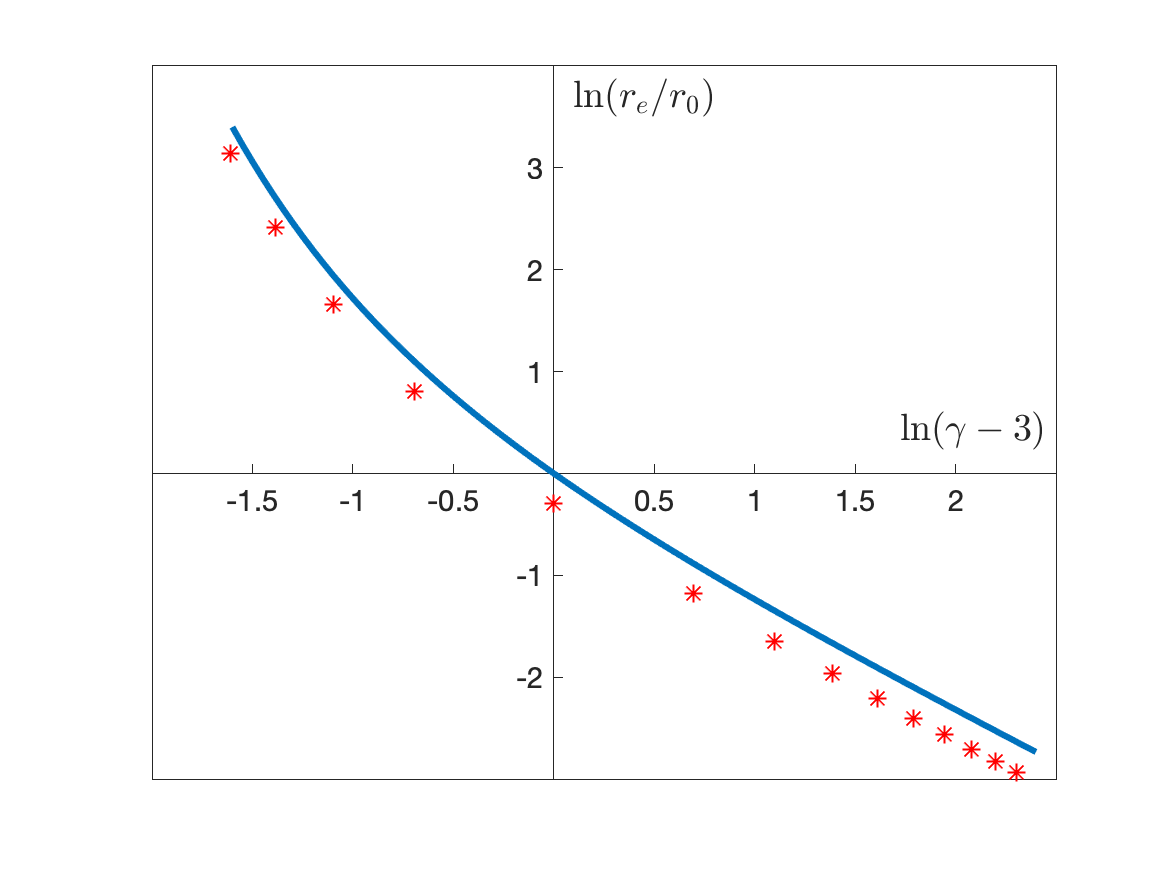} 
   \caption{The analytic expression for  $R_{1/2}/r_0$ compared with the numerical computation of $r_e/r_0$ for $\gamma$ between 3.2 and 13.
   }
   \label{fig:fp}
\end{figure}

Assuming a constant mass to luminosity ratio, the light distribution in the sky is given by integrating this density along lines of sight. Consequently, the equivalent radius $r_e$ is determined by the requirement
\begin{equation}
\int_0^{r_e} \int_{-\infty}^\infty  \frac{dz}{(r^2+z^2)(1+ \sqrt{r^2 +  z^2}/r_0)^{\gamma-2}} r dr = \frac{r_0}{\gamma-3}
\label{requivalent}
\end{equation}
We have the elementary bound $r_e < R_{1/2}$ from the fact that an infinite cylinder with radius $R_{1/2}$ contains in it a sphere of radius $R_{1/2}$. We might expect that $r_e \sim R_{1/2}$. This is indeed true, as illustrated by Fig.~\ref{fig:fp}, which plots the numerically obtained solutions of~\eqref{requivalent} overlayed with the curve in~\eqref{scale-length}. We see that $r_e < R_{1/2}$ but is close over the entire range $3.1 \leq \gamma \leq 15$ and this gives the relation
\begin{equation}
\frac{1}{\gamma-3} \approx \frac{\ln(1+ r_e/r_0)}{\ln 2}.
\label{powerlaws}
\end{equation}

%%%%%%%%%%%%%%%%%%%%%%%%%%%%%%%%%%%%%%%%%%%%%%%%
%
%		Fundamental Plane
%
%%%%%%%%%%%%%%%%%%%%%%%%%%%%%%%%%%%%%%%%%%%%%%%%

\subsection{The fundamental plane} \label{sec:FP}

Luminous/massive elliptical galaxies have a light distribution that is fit well by the de Vaucouleurs law \cite{deVaucouleurs1948Recherches}. A simple model for a (3d) mass distribution, giving projected light curves that are consistent with observations at the same level as de Vaucouleurs law, is the Jaffe profile $\rho_B(r) \sim r^{-2}$ for small $r$ and $\sim r^{-4}$ for large $r$ \cite{Jaffe1983Simple}. The Jaffe profile arises naturally in our framework and corresponds to $\gamma = 4$. Importantly, we get this profile from an isothermal (i.e. ``simple", isotropic) distribution function. Rather than pick a particular value of $\gamma$, we propose that Luminous/massive elliptical galaxies are (potentially) described by a range of possible values $\gamma \geq 4$ and that diffuse dwarf elliptical galaxies are described by $\gamma \approx 3$.

We now determine the relation between the equivalent radius $r_e$, the surface brightness $I_e \propto L/r_e^2$ and the velocity dispersion $\sigma^2$ where the luminosity $L \propto M_B$. From~\eqref{mass} we get
\begin{equation}
I_e \propto \frac{M_B}{r_e^2} \sim \frac{\sigma^2 r_0}{(\gamma-3) r_e^2} \sim  \frac{\sigma^2 r_0 \ln(1+r_e/r_0)}{ r_e^2}
\label{fp1}
\end{equation}
Although the logarithm cannot be uniformly approximated by a power law over the entire allowed range of $r_e/r_0$, we have $\ln(1+r_e/r_0) \approx r_e/r_0$ for $r_e \ll r_0$. In this situation, $r_0$ drops out of~\eqref{fp1} and we recover the result  $I_e \propto \sigma^2 r_e^{-1}$ suggested by the virial theorem. 

Although there isn't an universal power law relating $(\gamma-3)$ and $r_e/r_0$, for any given nominal value of $r_e/r_0$, or equivalently a given value of $\gamma$, we have a ``local" power law given by the exponent
$$
\alpha = \left. \xi \frac{d}{d\xi} \ln\left(\frac{\ln(1+\xi)}{\ln 2}\right)\right|_{\xi =  r_e/r_0} = \frac{r_e}{(r_e+r_0 ) \ln(1+r_e/r_0)}. 
$$ 
so that $\alpha$ varies from $\alpha = 1$ in the limit $\gamma \to \infty, r_e/r_0 \to 0$ to $\alpha = 0$ in the limit $\gamma \to 3, r_e/r_0 \to \infty$. In terms of the ``local" exponent $\alpha$, we get
$$
I_e \propto \frac{\sigma^2 r_0^{1-\alpha}}{r_e^{2-\alpha}}.
$$
Rearranging yields 
$$
r_e \sim \sigma^{\frac{2}{2-\alpha}}  I_e^{-\frac{1}{2-\alpha}} r_0^{\frac{1-\alpha}{2-\alpha}} 
$$ 
As we discussed above, there is some uncertainty in determining the appropriate values of $\alpha$. Nonetheless, we can draw the following qualitative conclusions:
\begin{enumerate}
\item The fundamental plane for diffuse galaxies ($\gamma \approx 3, \alpha \approx 0$) is distinct from the plane for luminous/massive elliptical galaxies ($\gamma \geq 4, \alpha \approx 1$) as borne out by observations \cite{Gudehus1991Systematic}.
\item The dependence on $r_0$ is very weak (an exponent between 0 and 1/2) and, in general, the relation is of the form $r_e \propto \sigma^{n_1} I_e^{-n_2}$ where $1 < n_1 < 2$ and $1/2 < n_2 < 1$. This is certainly the case for the (usual) fundamental plane $r_e \sim \sigma^{1.4} I_e^{-0.8}$ for luminous galaxies \cite{Djorgovski1987Fundamental}.
\end{enumerate}
As a final comment, our analysis in this section is for isothermal distributions, although our methods generalize directly and can be applied to more general ergodic distributions $f(\mathbf{x},\mathbf{v}) = f\left(\phi(\mathbf{x}) + \frac{|\mathbf{v}|^2}{2}\right)$. Also, it would be interesting to compare the predictions from our model with the corresponding results from MOND \cite{Cardone2011MOND-FP}.

%%%%%%%%%%%%%%%%%%%%%%%%%%%%%%%%%%%%%%%%%%%%%%%%
%
%		Disk+Bulge
%
%%%%%%%%%%%%%%%%%%%%%%%%%%%%%%%%%%%%%%%%%%%%%%%%

\section{Disk galaxies with bulges} \label{sec:HSB}

In this section we will construct self-consistent solutions of systems that contain both a pressure supported 3d component, i.e. a bulge, and a rotation supported thin disk. Such systems are ubiquitous and we argue that HSB galaxies, or indeed any system where the projected surface density is larger than $\Sigma^*$ {\em necessarily needs pressure support}. In this endeavor, we are guided by the insights from~\cite{Brada1995Exact} in the construction of disk galaxies with bulges in the context of MOND. In section~\ref{sec:RAR} we collect our results from the preceding sections to argue that our framework naturally leads to many of the observed galaxy scaling relations including the BTFR, the Faber-Jackson relation and the fundamental plane relation. In particular we highlight the various radial acceleration relations (RAR) that arise from our framework, and highlight the following testable prediction -- For purely rotation supported systems, the RAR has 2 branches for sufficiently small accelerations, in contrast to systems with pressure support where the RAR has a single monotonic branch.

To construct galaxy distribution functions for disk+bulge galaxies, we will exploit the fact that, to the extent that both (Lagrangian theories of) MOND \cite{Beckenstein1984does,Beckenstein2004TeVeS,Skordis2020RelMOND} and our framework give valid descriptions of galaxies, they should be related to each other, and there is potential for transferring results from one formulation to the other. In the initial formulation of MOND \cite{Milgrom_MOND_1983}, the Newtonian gravitational potential $\phi_N$ was only sourced by the baryonic matter density $\rho_B$, $\Delta \phi_N = -4 \pi G \rho_B$, while the dynamics was given by 
\begin{equation}
\ddot{\mathbf{x}} = -\mu\left(\frac{|\nabla \phi_N|}{a_0}\right) \nabla \phi_N
\label{MOND}
\end{equation}
for an appropriate transition function $\mu$ and a universal acceleration scale  $a_0 \sim 10^{-10} m/s^2$ \cite{Milgrom_MOND_1983,Famaey2012MOND}. 
This formulation of MOND is therefore predicated on the claim that $\ddot{\mathbf{x}}$ is {\em determined locally} as a function of the Newtonian gravitational acceleration $\nabla \phi_N$ sourced {\em purely by baryonic matter}. This idea has strong observational support in the radial acceleration relationship (RAR) \cite{McGaugh2016RAR,Lelli2017onelaw}, and we will discuss this further below.

Eq.~\eqref{MOND}, however, can only serve as an approximate formulation of the true dynamics since $\mu\left(\frac{|\nabla \phi_N|}{a_0}\right) \nabla \phi_N$ {\em is not, in general, a conservative force field}. The dynamics should be formulated through a Lagrangian having the right symmetries, so that the usual conservation laws of energy, momentum and angular momentum follow \cite{Beckenstein1984does}. On the other hand, if  $|\nabla \phi_N| = F(\phi_N)$, then  $\nabla \times \mu\left(\frac{|\nabla \phi_N|}{a_0}\right) \nabla \phi_N =0$. The approximate dynamics  $\ddot{\mathbf{x}} = -\mu\left(\frac{|\nabla \phi_N|}{a_0}\right) \nabla \phi_N$  are indeed conservative and {\em do represent the full dynamics of MOND in certain (Lagrangian) formulations} \cite{Brada1995Exact}. 

It is therefore interesting to study matter distributions for which the Newtonian potential satisfies  $|\nabla \phi_N| = F(\phi_N)$. This condition holds for matter distributions with a high degree of symmetry, for instance spherical solutions, but is not restricted to such distributions. Also, given a solution of $\Delta \phi_N = - 4 \pi G \rho$ satisfying  $|\nabla \phi_N| = F(\phi_N)$, we  construct a new potential through
$$
\tilde{\phi}(x,y,z) = \phi_N(x,y,z_0+|z|) = \begin{cases} \phi_N(x,y,z_0+z) & z \geq 0 \\ \phi_N(x,y,z_0 - z) & z \leq 0
\end{cases}
$$
$\tilde{\phi}$ is clearly continuous and satisfies $|\nabla \tilde{\phi}| = F(\tilde{\phi})$ for the same function $F$ and for all $z \neq 0$. We can compute the corresponding mass density as the Laplacian of the potential $\tilde{\rho} = -\frac{1}{4 \pi G} \Delta \tilde{\phi}$. 

The potential has a jump in the $z$-derivative across $z=0$, corresponding to a surface density (singular) component along the plane $z=0$. The density is given by
$$
\tilde{\rho}(x,y,z) = \rho(x,y,z_0+|z|) - \frac{1}{2 \pi G} \partial_z \phi_N(x,y,z_0) \delta(z)
$$
Since $\tilde{\rho}$ has a continuous and a singular component, with appropriate choices of the density/potential pair $(\rho,\phi_N)$ and the reflection plane $z=z_0$, we can obtain solutions corresponding to razor-thin disks with bulges, not only in Newtonian gravity as outlined above, but also in a Lagrangian formulation of MOND to yield 
$$
\ddot{\mathbf{x}} = \begin{cases} -\mu\left(\frac{|\nabla \tilde{\phi}|}{a_0}\right) \nabla \tilde{\phi} & z \neq 0 \\
-\mu\left(\frac{|\nabla \tilde{\phi}(x,y,0^+)|}{a_0}\right) (\partial_x \tilde{\phi}(x,y,0), \partial_y \tilde{\phi}(x,y,0),0) & z \neq 0
\end{cases}
$$
To ensure that the surface density be positive, it is necessary and sufficient that $\partial_z \phi_N(x,y,z_0) < 0$ for all $x,y$.

We will now extend these idea to our framework. In the weak field limit of our theory, there are two fundamental scalar fields the gravitational potential $\phi$ and the pattern phase $\psi$. The theory is invariant to shifts in $\phi$ and $\psi$, so the ``physical" fields are the gradients $\nabla \phi$ and $\nabla \psi$. We will consider the subclass of solutions that satisfy 
\begin{equation}
\nabla \psi \times \nabla \phi = 0,
\label{Brada-Milgrom}
\end{equation}
i.e. solutions for which the equipotentials $\phi = $ constant are identical with the phase surfaces $\psi =$ constant. This is the analog of the condition $|\nabla \phi_N| = F(\phi_N)$ in our setting. The motivation for considering such solutions, and the subsequent construction, comes from the work of Brada and Milgrom on similar ideas for constructing exact solutions in MOND \cite{Brada1995Exact}, as outlined above. We will therefore call solutions that satisfy the condition in Eq.~\eqref{Brada-Milgrom} the {\em Brada-Milgrom solutions}. 

Among these solutions are the time independent radial solutions $\phi = \phi(R), \psi = \psi(R)$, corresponding to  baryonic densities $\rho = \rho_B(R)$, that we've constructed in Sec.~\ref{sec:spherical}. The phase for a radial solution is a monotonic function of the radius, so that, with the normalization $\psi(0) = 0$, $\psi$ is uniquely determined by $R$ and vice-versa. 

We can now build a solution with azimuthal symmetry starting from the spherical solutions $\phi_0(R)$ and $\psi_0(R)$. In cartesian coordinates $(x,y,z)$ define $R_{\pm} = \sqrt{x^2+y^2 + (z_0\pm z)^2}$, and in all space, let $\phi(x,y,z) = \phi_0(R_+)$ for $z \geq 0$ and $\phi(x,y,z) = \phi_0(R_-)$ for $z \leq 0$, and similarly for $\psi$. $\phi$ and $\psi$ are continuous functions, but there are, in general, jumps in $\nabla \phi$ and $\nabla \psi$ across the plane $z=0$. There are two jump conditions, one from the Poisson equation $\Delta \phi = -4 \pi G(\rho_B+\rho_P)$ and the second from the Schr\"odinger equation $(-\Delta + W)e^{-\psi} = -k_0^2 e^{-\psi}$. 

Note that, the (inferred) potential  for the Schr\"odinger equation determines $\sqrt{\tilde{\rho}} \simeq \sqrt{\Sigma_B/w}$ (see Eq.~\eqref{disk-k0} in Appendix B) where $w$ is the width of the PGB boundary layer as illustrated in Fig.~\ref{fig:PGB}. Likewise, from matching the jump in $\partial_z \phi$ with the mass density on the disk given in Eq.~\eqref{disk-energy} we get 
\begin{align}
-\frac{1}{2 \pi G}  \phi_z(x,y,z_0) & \simeq \frac{8 \Sigma^*}{3} \sin^3 \theta(s) + \Sigma_B(s) V(k_0^2 \cos^2\theta(s)) \nonumber \\
& \approx \Sigma_B(s) \left[\frac{2 V_0}{3} - \ln\left(1 - \left(\frac{V_0 \Sigma_B(s)}{4 \Sigma^*}\right)^{2/3} \right)\right]
\label{disk-jump}
\end{align}
where the last line follows from~\eqref{galaxy} and $s = \sqrt{x^2+y^2}$.  

We again see that $\Sigma_B(s) \leq 4 \Sigma^*/V_0$ and, as we discussed earlier, the purely ``cold-disk" component of the any galaxy can only support a maximum surface density of the order of $\Sigma^*$. Conversely, in situations where the effective surface density is larger than $\Sigma^*$, the system cannot be entirely rotation-supported, and a fraction of the mass has to be in a pressure supported bulge.

We are now in a position to construct self-consistent disk+bulge galaxy DFs and thus describe HSB galaxies. The DF consists of {\em two components}, $f = f_{\text{bulge}} + f_{\text{disk}}$, where, in terms of cylidrical coordinates $(r,z)$, we have
\begin{align}
\tilde{\phi}(r,z) & = \phi(r,|z|+z_0) \nonumber \\
f_{\text{bulge}} & = \frac{\rho_0}{(2 \pi \sigma^2)^{3/2}}\exp\left(-\frac{1}{\sigma^2}\left(\frac{\mathbf{v}^2}{2} + \tilde{\phi}(r,z) \right) \right) \nonumber \\
 \quad \Rightarrow \quad \rho_{\text{bulge}}(r,z) & = \rho_0 \exp\left(-\frac{\tilde{\phi}(r,z)}{\sigma^2}\right)
\label{fbulge}
\end{align}
for the bulge component. Note that, for the isothermal DF, or more generally for any ergodic DF $f = f\left(\frac{\mathbf{v}^2}{2}+\tilde{\phi}\right)$, the DF satisfies the collisionless Boltzmann equation in all of phase space despite the jump in $\nabla \tilde{\phi}$ across $z=0$ because there velocity distribution is isotropic everywhere, and the distribution on the velocity variables is smooth at $z=0$. 

For the disk component, we determine $\Sigma_B$ by solving \eqref{disk-jump}. Since $E$, the total energy, and $L_z$, the $z$-component of the angular momentum are conserved, any function of the form $f=f(L_z,E)$ is a solution of the collisionless Boltzmann equation. Following the discussion in Dehnen \cite{Dehnen1999Approximating,Dehnen1999Simple},  we define
\begin{align*}
L_c(r) & = r \left[ -r \partial_r \tilde{\phi}(r,0) \right]^{1/2}, \nonumber \\
E_c[r] & = \frac{L_c(r)^2}{2r^2} + \tilde{\phi}(r), \nonumber \\
\gamma & = \frac{2 \Omega}{\kappa},   
\end{align*}
where $L_c$ is the specific angular momentum, $E_c$ is the energy and  $\Omega$ and $\kappa$ are respectively the angular and epicyclic frequencies (see Sec.~\ref{sec:stability} for the relevant definitions), for circular orbits in plane $z=0$ in the potential $\tilde{\phi}(r,z)$. From the discussion in Sec.~\ref{sec:spherical}, we have $\tilde{\phi}(r,0) = \phi\left(\sqrt{r^2+z_0^2}\right) \sim \ln r$ for $r \gg z_0$. Consequently, $L_c(0) = 0$ and $L_c(r)$ grows like $r$ for large $r$ so that, for any given value $L > 0$, there is a solution $r = R_L$ to the equation $L_C(r) = L$. In terms of this radius $R_L$, we have the {\em cold-disk distribution function}
\begin{equation}
f_{\text{disk}}(L,E)  = \frac{\gamma(R_L)\Sigma_B(R_L)}{2 \pi} \delta\left(E-E_c(R_L)\right)
\label{cold-disk-DF}
\end{equation} 
which includes effects, at the lowest order, due to deviations from circular orbits \cite{Dehnen1999Approximating}. This DF can be ``warmed up" following the prescription in \cite{Dehnen1999Simple}.

We illustrate this procedure by starting with a spherical galaxy with $\gamma = 4$. From Eq.~\eqref{mass}, the total baryonic mass of a spherical galaxy with $\gamma =4$ is $M_B = \frac{2 \sigma^2}{G k_0}$ and the fraction of this mass in the resulting bulge is given by
$$
\frac{M_{\text{bulge}}}{M_B} = \int_{z_0}^\infty \int_0^\infty \frac{k_0 r dr dz}{(r^2+z^2) (1+k_0\sqrt{r^2+z^2})^2} = 1 - k_0 z_0 \ln \left(1+\frac{1}{k_0 z_0}\right)
$$ 

\begin{figure}[htbp] %  figure placement: here, top, bottom, or page
   \centering
   \includegraphics[width=0.7 \textwidth]{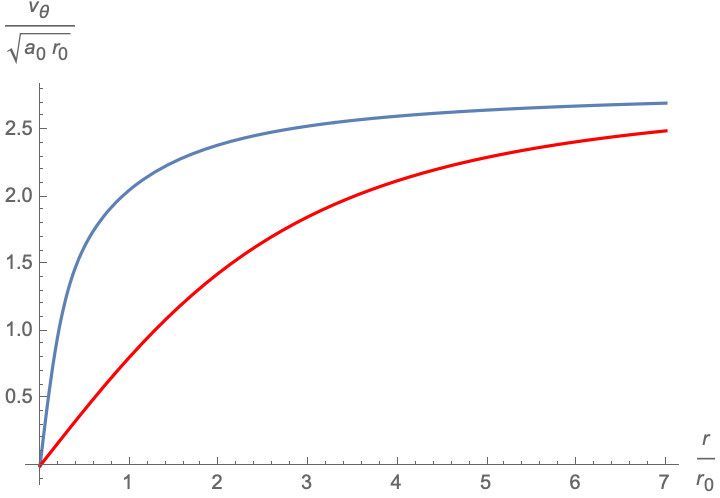} 
   \caption{The rotation curves for the galaxies given by the Brada-Milgrom solutions with $\gamma=4$ and $z_0 = \frac{r_0}{3}$ (blue curve) and $z_0 = 3 r_0$ (red curve). The blue curve is typical of HSB galaxies that rise quickly and then saturate, while the red curve is typical of LSB galaxies that continue ot rise, albeit slowly, over many scale lengths.
   }
   \label{fig:disks}
\end{figure}

For $\gamma = 4$ and $z_0 = \frac{1}{3} r_0$, we get $M_{\text{bulge}} \approx 0.53 M_B$ so that about half the mass of the galaxy is in the bulge in this case. By way of contrast, for $z_0 = 3 r_0$, we have $M_{\text{disk}} \approx 0.13 M_B$ so the bulk of the baryonic mass is in the disk.  Figure~\ref{fig:disks} show the rotation curves for the resulting disk+bulge galaxies.

%%%%%%%%%%%%%%%%%%%%%%%%%%%%%%%%%%%%%%%%%%%%%%%%
%
%		RAR
%
%%%%%%%%%%%%%%%%%%%%%%%%%%%%%%%%%%%%%%%%%%%%%%%%

\subsection{Galaxy scaling relations from pattern dark matter} \label{sec:RAR}

The self-consistent solutions of spherical galaxies are determined by two quantities ``external" quantities, the total baryonic mass $M_B$ and the velocity dispersion $\sigma^2$, or equivalently, the total kinetic energy $\frac{3}{2} M_B \sigma^2$. The dimensional parameters in our theory are Newton's constant $G$ and the surface density scale $\Sigma^*$ or equivalently the acceleration scale $a_0$. All the other quantities in our theory, for example $k_0,r_0$ or $r_e$, emerge from the dynamics. It follows from dimensional analysis that the solutions are characterized by a single dimensionless parameter. A natural choice for this parameter is the ``Faber-Jackson" combination $G M_B a_0 \sigma^{-4}$. 

As we see from~\eqref{mass}, this is equivalent to choosing $\gamma$ as the unique (nondimensional) parameter governing the dynamical equilibrium of an isothermal, spherical galaxy. In particular, dimensional analysis implies a relation of the form $\frac{g_{\text{obs}}}{a_0} = f\left(\frac{g_{\text{bar}}}{a_0},\gamma\right)$. From Eqs. \eqref{scaling_regime}~and~\eqref{massltR} we get
$$
g_{\text{bar}}(R) = \frac{G M_B}{R^2} = \frac{8 a_0}{\sigma^2 \gamma k_0 R} \frac{2 \sigma^2 r_0}{R (\gamma-3)}\left[1-\frac{1}{(1+R/r_0)^{\gamma-3}}\right].
$$
In conjunction with~\eqref{spherical-soln}, we get that $g_{\text{bar}}/a_0$ is a function of the dimensionless combination $\xi = k_0 R$ that also depends on $\gamma$. Likewise, from Eqs.~\eqref{eq:halo_mass}~and the identification $a_0 = 2 \pi G \Sigma^*$ we get
$$
g_{P}(R) = \frac{G M_P}{R^2} = \frac{4 a_0}{k_0 R} \left(2 - \frac{\sqrt{2}}{k_0 R} \arctan\left(\sqrt{2}k_0 R\right) \right),
$$
and that $g_P/a_0$ is only a function of the same combination $k_0 R$. 

For small $R$, $g_{\text{obs}} \approx g_{\text{bar}} \gg g_P$ while for large $R$, we have  $g_{\text{obs}} \approx  g_P \sim \sqrt{g_{\text{bar}} \frac{4\eta_c^2 a_0}{\pi}}$, which is consistent with the MOND rule for small baryonic accelerations, $g_{\text{obs}} \approx \sqrt{g_{\text{bar}} g_{\dag}}$, with the definition $g_{\text{dag}} = 8 \eta_c^2 G \Sigma^*$. We emphasize that this rule was not baked into the effective Largangian in~\eqref{Lagrangian},  but rather, is a dynamical consequence of the self-organization of the baryonic distribution and the associated pattern field. In particular, $\eta_c$ can depend on the parameter $\gamma$ describing the underlying galaxy DF, so we don't have just one RAR, i.e. an unique relation $g_{\text{obs}} =  a_0 f\left(\frac{g_{\text{bar}}}{a_0}\right)$, but rather an entire family of such functions that depend on the underlying galaxy DF (in our discussion this is through the parameter $\gamma$), although they all interpolate between the same limit (scaling) behaviors for $\frac{g_{\text{bar}}}{a_0} \gg1$ and $\frac{g_{\text{bar}}}{a_0} \ll 1$.

Finally, our self consistent solutions of disk + bulge galaxies are generated starting from spherical isothermal solutions, and using the Brada-Milgrom procedure. Consequently they are determined by 3 parameters, the total baryonic mass $M_B$ and the parameter $\gamma$ of the underlying spherical solution, and the relative offset $z_0/r_0$ or equivalently the mass ratio $\beta = M_{\text{disk}}/M_B$ of the galaxy. The same dimensional considerations again apply, and we get that $\frac{g_{\text{obs}}}{a_0} = f\left(\frac{g_{\text{bar}}}{a_0},\gamma, \beta \right)$

\begin{figure} %  figure placement: here, top, bottom, or page
  \centering
  \includegraphics[width=0.8 \textwidth]{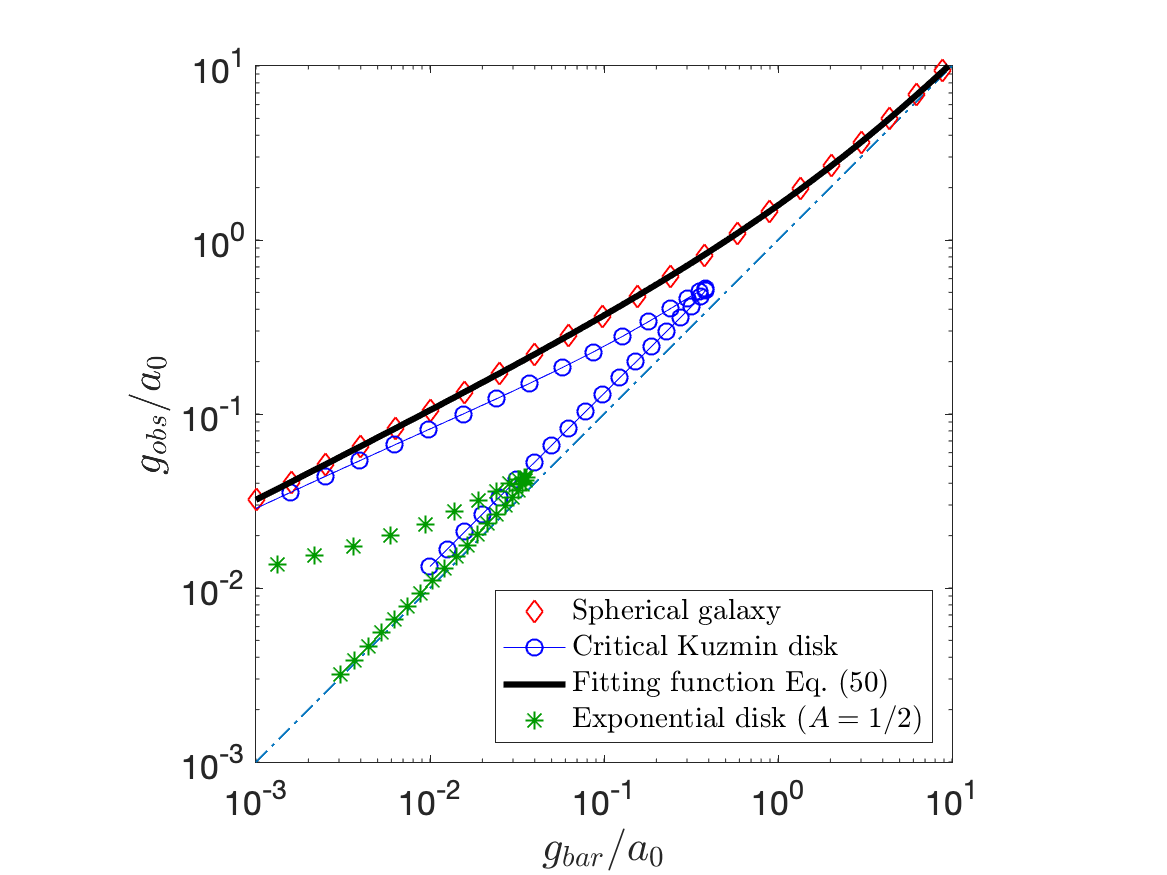} 
  \caption{The radial acceleration relation (RAR) in our model for various distributions of matter.   $g_{\text{obs}}$ and $g_{\text{bar}}$ are  the total gravitational acceleration and the baryonic contribution respectively.  Compare Fig.~1 in Ref.~\protect{\cite{DiPaolo2019RAR}}. }
  \label{fig:rar}
\end{figure}

In Fig.~\ref{fig:rar}, we plot the RAR for our theory applied to various matter distributions ($V_0$ is set to 4) and compare with the fit
\begin{equation}
 g_{\text{obs}} = \frac{g_{\text{bar}}}{1- e^{-\sqrt{g_{\text{bar}}/g_{\dag}}}} 
 \label{eq:fitting}
\end{equation} 
for the choice $g_{\dag} = a_0$ \cite{Lelli2017onelaw}, where we have set $4\eta_c^2/\pi = 1$ for the case of the spherical galaxy. The RAR for the spherical galaxy agrees with the fitting function over the entire range (4 decades) in $g_{bar}/a_0$. Note that, for rotation supported systems, the resulting RAR has two branches. In the absence of a bulge/central mass,  the baryonic contribution to the acceleration $g_{\text{bar}}$ has a peak value $g_{\max}$ at a few scale-lengths. For $g < g_{\max}$ there are two values of $r$, one on either side of the peak, with $g_{\text{bar}}(r) = g$ and (generically) {\em different} values of $g_{\text{obs}}$, giving two branches that meet at the peak acceleration. Our theory therefore gives two branches for the RAR in agreement with recent observations for dwarf disk and LSB galaxies~\cite{DiPaolo2019RAR}. In contrast, pressure supported systems have a monotonic decrease in  $g_{\text{bar}}$ and have a single branch, although the relation is not universal or intrinsic; rather it is an emergent characteristic  of dynamically self-organized LSB galaxies.

%%%%%%%%%%%%%%%%%%%%%%%%%%%%%%%%%%%%%%%%%%%%%%%%
%
%		Discussion
%
%%%%%%%%%%%%%%%%%%%%%%%%%%%%%%%%%%%%%%%%%%%%%%%%

\section{Discussion} \label{sec:discuss}

We have proposed an effective Lagrangian field theory for dark matter built on ideas from pattern formation. In this process we have introduced an additional ``dark field" $\psi$ that plays the role of DM. In our theory, no structures are formed on scales smaller than $k_0^{-1}$  resulting 
in cored DM halos in contrast to the cuspy halos formed by CDM. 

 We would like the reader to treat our framework as a mathematical model and, although it is motivated by the physical picture we have drawn, it is not necessarily beholden to it. It is well known in the physical sciences that results that have the correct scaling properties do not verify the truth of the physical picture which motivated the model. We must beware of ``looks like" science. A great example is again drawn from convection and to the original observation of Rayleigh-Benard convection in a horizontal  layer of fluid heated from below with its top surface free. The resulting hexagonal pattern, hexagonal rather than roll-like because the system is not symmetric about the mid-layer and thus the up-down symmetry is broken, behavior consistent with hexagons, had all the hallmarks of what we expect from convection driven by adverse temperature gradients. But it was not. The driving mechanism was surface tension gradient effects arising from different temperatures on the free surface, the so called Marangoni effect. Likewise, in interpreting the successes that our model displays, we must be open to the possibility that there may be other mechanisms beyond the one based on the energy stored in pattern defects which may give rise to the same outcomes and have similar scaling laws. 
 
In other words, our theory is based on ``universal" equations and can thus describe a variety of physical mechanisms. For instance, BEC DM \cite{Sin1994BEC}, Self-interacting DM \cite{Spergel2000SIDM}, Superfluid DM \cite{Berezhiani2015theory} and Fuzzy DM \cite{Hu2000fuzzy}  all have associated scalar fields that could  play the role of $\psi$. Our favored interpretation is that $\psi$ is the order parameter for a broken translational symmetry, with a characteristic scale $k_0^{-1}$ determined by the distribution of baryons. For disk galaxies, we identify $k_0$ with the (average) maximally unstable wavenumber in  the stability analysis that leads to Toomre's criterion \cite{Toomre1964Gravitational}, thereby directly relating the `baryonic instability' of a rotating disk to the pattern instability that produces the `dark halo' in our framework. More generally, we determine $k_0$ self-consistently from a nonlinear eigenvalue problem, that is motivated by analogies with instabilities in chemical patterns \cite{Kopell1981Target}.

 Ours is an effective, long wave theory, applicable on scales $\gtrsim k_0^{-1}$, rather than a fundamental theory, since $k_0$ in~\eqref{Lagrangian} is a nonlocal object that depends on the distribution of baryons. This nonlocality is to be expected and is indeed unavoidable for an effective theory that is consistent with the BTFR \cite{Deffayet2011nonlocal}. 
  
There are two distinct sources for the new effects arising in our theory. First, the curvature of the phase surfaces contributes an additional energy (mass), consistent with cored halos. The resulting gravitational acceleration dominates the baryonic contribution $g_{\text{bar}}$ at large distances and flattens the rotation curves. Second, for disk galaxies, the phase surfaces are spheroidal rather than spherical and thus generate a phase grain boundary on the galactic plane. This additional source yields the third relation in Eq.~\eqref{galaxy}, linking $\theta(s)$, the angle between the phase surfaces and the galactic plane, to the density of the disk, $\Sigma(s)$,  providing a natural explanation for the disk-halo connection in galaxies \cite{Sancisi2004visible,DiPaolo2019universal}.  One consequence of this connection is that exponential disks correspond to a constant curvature for the phase surfaces on the galactic disk $z=0$, and thus to a constant `halo' surface density \cite{Donato2009constant}.

We have only investigated quasi-steady rotation supported systems, and it is of great interest to understand dynamical effects that come from time varying pattern fields. There are multiple ramifications of this issue. Firstly,  the Lorentz invariant pattern Lagrangian in~\eqref{Lagrangian} has higher order time derivatives and is therefore subject to the Ostrogradsky instability. This motivates the need to study theories that break Lorentz invariance, and only have first order time derivatives in a preferred time-slicing, as in  Lifschitz-Horava theories. Additionally, the arguments for the dynamical origin of a preferred wave-number $k_0$ in chemical waves, that were the inspiration for our argument in Sec.~\ref{sec:dynamical}, pertain to time varying patterns, and the dynamical selection of $k_0$ is only evident on times scales $\tau \sim (\omega_1-\omega_0)^{-1}$ where $\omega_0$ is the frequency associated with the ground state and $\omega_1$ is the frequency associated with the first excited state of the relevant Schr\"odinger operator. In particular, for chemical waves, the existence of this time scale allows for transient behavior with multiple ``localized" defects, each with its own preferred wavelength, transitioning into a single global pattern eventually. We expect that a similar effect will play a role for interacting galaxies including satellites and colliding galaxies or clusters. Indeed, in analogy with defects in patterns, Galaxy collisions are unsteady phenomena that can potentially generate new defects. In the context of our theory, they are accompanied by (potentially) large disruptions of existing quasi-steady patterns and changes in $k_0$ on a short time scale. Further work is needed to understand these processes.

Our theory does predict novel effects that we are yet to investigate. For instance, if the jump in $\nabla{\tilde{\psi}}$ at the galactic plane is too sharp, the pattern is unstable to the generation of codimension 2 defects called {\em disclinations} \cite{newell1996defects,patt_defects}. These defects take the form of spirals, for a traveling wave of disclinations in 3-space. Spiral disclinations create a periodic forcing in the gravitational potential which can excite and sustain density waves in the baryonic matter that {\em will be correlated} with the disclinations in the underlying pattern. This is a fully nonlinear, far from threshold phenomenon and might be relevant to understanding the surprising nonlinear stability of spiral arms over many rotation periods \cite{Bertin2014dyga.book}. 

In ongoing work we are  (1) Addressing the Ostrogradski instability for the action in Eq.~\eqref{Lagrangian}, by breaking Lorentz invariance as in Ho\v{r}ava gravity \cite{Horava2009quantum,Blas_Consistent_2010}, to get a Lagrangian only involving first order time derivatives in a preferred slicing, (2) Varying $k_0$ in space for studying clusters of galaxies and compare the effective dark matter in our model with what is inferred from gravitational lensing, and (3) Evaluating the consequences  of the potential instability of the PGB when the angle $\theta(s)$ becomes too sharp  \cite{newell1996defects}. 

Our model allows us to compute the rotation curves of exponential disk galaxies. The resulting phase contours for exponential disks  are (approximately) spherical caps $\psi \approx (r^2 + (z+z_0)^2)^{1/2}$, as illustrated in Fig.~\ref{construction}, implying that the Kuzmin disk solutions approximate the ``dark halos" of disk galaxies \cite{Brada1995Exact}. In other contexts, our model allows us to compute the galaxy distribution functions for systems with pressure support, including spherical and disk+bulge galaxies. Our model suggests a possible dynamical origin for the MOND rule $\ddot{\mathbf{x}} = -\mu\left(\frac{|\nabla \phi|}{a_0}\right) \nabla \phi$ \cite{Milgrom_MOND_1983} where $\phi$ is the Newtonian potential sourced purely by baryonic matter $\Delta \phi = 4 \pi G \rho$. In this sense, the MOND rule is akin to a galaxy scaling relation, i.e. a manifestation of an underlying dynamical self-organization, rather than a fundamental law of nature, and as such, one {\em should not expect} a single universal transition function $\mu$ to {\em describe the dynamics of galaxies in all situations}.

Galaxy formation is a complex process, and involves a great many effects \cite{Dalcanton1997Formation,Wechsler2018Connection} not included in our simple model. In a cosmological context galaxies are ``nonlinear", with the implication sometimes being that theorists can build models that describe the very largest scales of the universe, and not be too concerned with tensions between theories and observations, or ``unexplained" regularities/scaling laws, on small ``nonlinear" scales \cite{Famaey2012MOND}. We disagree with this point of view.
  We contend that the robust scaling relations satisfied by galaxies are not ``accidental" and require robust explanations. Our model offers conceptual insight into these relations by demonstrating how a generic mechanism for coupling the dark field $\psi$, through its defects, to the baryonic density $\rho_B$ leads to self-organization.% 
  
Additionally, our model is a useful technical tool. The fact that galaxies are non-relativistic implies there is a well-defined $c \to \infty$ limit for our theory, that only involves three dimensional quantities $G,\Sigma^*$ and $k_0$ (or equivalently $M_B$). Assuming that the {\em `deep MOND'} limit \cite{Milgrom2009MOND} $M_B \to 0$ or equivalently $a_0 = 2 \pi G \Sigma^* \to \infty$ is also well defined, we reduce to two relevant dimensional quantities.  Dimensional analysis then {\em requires} the existence of a space-time scale invariance in this limit \cite{Milgrom2009MOND}. As we discuss elsewhere, we can build on this idea and embed our model~\eqref{Lagrangian} into a Renormalization group (RG) through scaling transformations for the quantities in~\eqref{galaxy}. Under the RG flow, we expect that the evolute $\gamma$ will degenerate to a single point, and the one parameter family of critical Kuzmin disks will be the corresponding fixed points. The critical Kuzmin disks therefore ``shepherd" the behavior of rotation supported disk galaxies. In particular, scaling relations for the Kuzmin disks, like the RAR we obtain in~\eqref{kuzmin-pot}, {\em ought to}  hold approximately for general disk galaxies, as in Fig.~\ref{fig:rar}.

Observed galaxy scaling relations suggest an underlying {\em universality} \cite{Kadanoff1990scaling} in the dynamical self-organization of galaxies, which in turn justifies our use of simplified physical models, with relatively few ingredients, in our investigation of self-organization in galaxies. Our results underscore the need for going beyond CDM and incorporating physical processes, including baryon coupling \cite{Famaey2020BIDM} and self-interaction \cite{Spergel2000SIDM}, in order to explain  observed dark matter phenomenology on galactic scales \cite{Navarro2017MDARfull,Dutton2019origin}. To the extent our `universal' model reproduces observations, it constrains theories with  baryonic feedback since  Eq.~\eqref{Lagrangian} should emerge as a limit theory in the appropriate scaling regime. Beyond discriminating among CDM + feedback theories, our model goes further in suggesting that  `dark matter' can arise as a collective, emergent phenomenon ({\em cf.} \cite{verlinde2017}) and not only as an yet undiscovered particle. The viability of this idea merits further study from an astrophysical viewpoint {\em and} from the viewpoint of complex systems/pattern formation.

\section*{Appendix A: The stress tensor for pattern dark matter}

\setcounter{equation}{0}
\renewcommand{\theequation}{A.\arabic{equation}}

We now follows the discussion in \cite{Newell2019pattern} and compute the Stress-Energy-Momentum $T^{\alpha \beta}$ corresponding to this solution using the Einstein-Hilbert prescription $T^{\alpha \beta} = \frac{2}{\sqrt{-g}} \frac{\delta \tilde{E}}{\delta g_{\alpha \beta}}$  where $\tilde{E}$ is the appropriate Lagrangian in curved space-time and $g = \mathrm{det}[g_{\alpha \beta}]$.  The minimally coupled \cite{MTW} pattern action $\mathcal{S}_P$ is: 
\begin{equation}
\mathcal{S}_P = \frac{\Sigma^*_0 c^2}{k_0^3}   \int \left\{(\nabla^\mu \psi \nabla_\mu \psi-k_0^2)^2 +  (\nabla^\mu \nabla_\mu \psi )^2\right\} \sqrt{-g} \ d^4x,
\label{curved}
\end{equation}
where the metric $g_{\alpha \beta}$ has signature $(- \,+\, +\, +)$, $\nabla_\mu$ is the corresponding covariant derivative. 
To obtain the stress tensor in a (background) flat space time, it suffices to consider variations $g_{\alpha \beta} = \eta_{\alpha \beta} + t \rho_{\alpha \beta}$ and compute all the variations to first order in $t$. To this end, we record the following relations for the inverse metric $g^{\gamma \delta}$, the Christoffel symbols $\Gamma^\sigma_{\gamma \delta}$ and the quantities that appear in $\mathcal{S}_P$:
\begin{align*}
g^{\gamma \delta} & = \eta^{\gamma \delta} - t \, \eta^{\gamma \alpha}\eta^{\delta \beta} \rho_{\alpha \beta} + O(t^2) \\
\Gamma^\sigma_{\gamma \delta} & = \frac{t}{2}\, \eta^{\sigma \xi} \left[ \frac{\partial \rho_{\gamma \xi}}{\partial x^\delta} +  \frac{\partial \rho_{\delta \xi}}{\partial x^\gamma} - \frac{\partial \rho_{\gamma \delta}}{\partial x^\xi}\right]+ O(t^2) \\
\sqrt{-g} & = 1 + \frac{t}{2} \eta^{\alpha \beta} \rho_{\alpha \beta} + O(t^2) \\
\nabla^\mu \psi \nabla_\mu \psi & =  \eta^{\gamma \delta} \partial_\gamma \psi \partial_\delta \psi  - t \, \eta^{\gamma \alpha}\eta^{\delta \beta} \rho_{\alpha \beta} \partial_\gamma \psi \partial_\delta \psi + O(t^2) \\
g^{\gamma \delta} \nabla_\gamma \nabla_\delta \psi & = \Box \, \psi - t \, \eta^{\gamma \alpha}\eta^{\delta \beta} \rho_{\alpha \beta}  \partial_{\gamma} \partial_{\delta} \psi - \frac{t}{2}\, \eta^{\gamma \delta} \eta^{\sigma \xi} \left[ \frac{\partial \rho_{\gamma \xi}}{\partial x^\delta} +  \frac{\partial \rho_{\delta \xi}}{\partial x^\gamma} - \frac{\partial \rho_{\gamma \delta}}{\partial x^\xi}\right] \partial_\sigma \psi + O(t^2)
\end{align*}
The Einstein-Hilbert stress tensor is now given by
$$
\left. \frac{d}{dt}\mathcal{S}_P\right|_{t=0} = \frac{1}{2} \int T^{\alpha \beta} \rho_{\alpha \beta} \,d^4x + \text{ boundary terms}.
$$
A straightforward but somewhat lengthy calculation now yields
\begin{align}
T^{\alpha \beta}  =  \frac{\Sigma^* c^2}{k_0^3} & \left\{   -4(\eta^{\mu \nu} \partial_\mu \psi \partial_\nu \psi-k_0^2) \eta^{\sigma \alpha}\eta^{\tau \beta} \partial_\sigma \psi \partial_\tau \psi  - 4\, \Box \psi\, \eta^{\sigma \alpha}\eta^{\tau \beta} \partial_{\sigma} \partial_{\tau} \psi  \right. \nonumber \\
& + 2 (\eta^{\alpha \tau} \eta^{ \beta \sigma} + \eta^{\beta \tau} \eta^{\alpha \sigma} - \eta^{\alpha \beta} \eta^{\sigma \tau} ) \partial_\tau (\Box \psi \, \partial_\sigma \psi) \nonumber \\
& + \left. \eta^{\alpha \beta} \left[(\eta^{\mu \nu} \partial_\mu \psi \partial_\nu \psi-k_0^2)^2 + (\Box \psi)^2\right] \right\}
\label{stress-tensor}
\end{align}
We can now express the energy density $T^{\alpha \beta}$ for the (stationary) phase field $\psi(R)$ with respect to a normalized basis $\{\mathbf{e}_t, \mathbf{e}_R, \mathbf{e}_\theta, \mathbf{e}_\phi\}$ induced by (spatial) spherical polar coordinates $(r,\theta,\phi)$.

We decompose the stress tensor into two pieces, $T_s$ coming from the ``stretching energy"  with density $(|\nabla \psi|^2 - c^{-2} (\partial_t \psi)^2 - k_0^2))^{2}$ and $T_b$ from the bending energy $(\Box \, \psi)^2$. We compute these quantities for a stationary, radial solution $\psi = \psi(r)$ to get
\begin{equation}
T_s \sim  \frac{\Sigma^*c^2}{k_0^3} \begin{pmatrix}  \tau_1 & 0 & 0 & 0 \\
0 & \tau_2 &
  0 & 0 \\
0 & 0 &-\tau_1 & 0 \\
0 & 0 & 0 & -\tau_1 \end{pmatrix} \qquad
T_b \sim   \frac{\Sigma^*c^2}{k_0^3} \begin{pmatrix}  \tau_3 & 0 & 0 & 0 \\
0 & \tau_4 &
  0 & 0 \\
0 & 0 &-\tau_3 & 0 \\
0 & 0 & 0 & -\tau_3 \end{pmatrix}
\label{stress}
\end{equation} 
where 
\begin{align}
\tau_1 & = (\psi'(R)^2-k_0^2)^2 \nonumber \\
\tau_2 & = (3 \psi'(R)^2 +k_0^2)(\psi'(R)^2- k_0^2) \nonumber \\
\tau_3 & = -\psi ''(R)^2-\frac{2 \psi '(R) \left(R \psi'''(R)+4 \psi ''(R)\right)}{R} \nonumber \\
\tau_4 & = \frac{8 \psi '(R)^2}{R^2}+\psi ''(R)^2-2 \psi'''(R) \psi '(R)
\label{geometric}
\end{align}

We remark on the expected structure of $T_s$ and $T_b$, {\em viz.} the off-diagonal stresses should be zero from time-reversal and spherical symmetries of the solution $\psi$, and further $T^{00} = -T^{\theta \theta} = - T^{\phi \phi}$ since the metric has signature  $(- \,+\, +\, +)$, and $\psi$ is independent of $t, \theta$ and $\phi$. Finally, the quantities $\tau_1, \tau_2, \tau_3$ and $\tau_4$ are constrained by the conservation of Energy-Momentum $\nabla_\mu T^{\mu \nu} = 0$. A calculation shows that, as expected,  these 4 conditions reduce to just one constraint on $\psi(r)$, namely that $\psi$ should satisfy the Euler-Lagrange equation~\eqref{eleqn}. 

The stress tensor associated with the field $\psi$ will act as source for the curvature of space-time as we discuss in the body of this paper.

\section*{Appendix B: Bound states for Gaussian wells.}

\setcounter{equation}{0}
\renewcommand{\theequation}{B.\arabic{equation}}

Following the discussion in Sec.~\ref{sec:dynamical}, we now estimate the ground state energies for spherical and oblate Gaussian wells, to determine the appropriate choices for $k_0$ for spherical (i.e. elliptic) and disk galaxies.

The 1d Gaussian well is given by a potential $V(x) = - V_0 e^{-\frac{x^2}{2 a^2}}$. Close to the minimum, this is approximated by the Harmonic potential $-V_0 + \frac{V_0 x^2}{2 a^2}$ suggesting that an estimate can be obtained by using the variational method with trial functions given by eigenfunctions for the Harmonic oscillator \cite{Nandi2010Quantum}. With the normalized wavefunction $\psi_0 = \left(\frac{\beta}{\pi}\right)^{1/4} e^{-\beta x^2/2}$, we have, from the variational principle \cite{griffithsQuantum}, 
$$
\int \left( \psi_0'^2 + V(x) \psi_0^2 \right) dx = \frac{\beta}{2} - V_0 \sqrt{\frac{2 a^2 \beta}{1+2a^2\beta}} \geq -k_0^2 
$$
where $-k_0^2$ is the ground state energy for the operator $-\Delta + V$.  For small $\beta > 0$, the bound is $\frac{\beta}{2} - V_0 a \sqrt{2 \beta}$ which can be made negative independent of how small $V_0$ is, i.e. we always have bound states. The optimal $\beta$ minimizes the bound, and is given by
$$
\frac{1}{2} =  V_0 \sqrt{\frac{2 a^2 \beta}{1+2a^2\beta}} \left[\frac{1}{\beta} - \frac{2a^2}{1+2a^2\beta} \right] = a^2 V_0 \sqrt{\frac{2}{a^2\beta(1+2 a^2 \beta)^3}}
$$
It is clear that the optimal $\beta$ is in the ``scaling form" $\beta = a^{-2} F(a^2 V_0)$. Considering the regimes $a^2 V_0 \gg 1$ and $a^2 V_0 \ll 1$ separately we get
$$
a^2 \beta \approx \begin{cases}  8(a^2V_0)^2  & a^2 V_0 \ll 1 \\  \sqrt{a^2V_0}  & a^2 V_0 \gg 1 \end{cases}
$$
In the regime $a^2 V_0 \gg 1$ we get the ``deep potential" limit
\begin{equation}
k_0^2 \gtrsim V_0 - \frac{\sqrt{V_0}}{2a}, 
\label{deep-limit}
\end{equation}
corresponding to the Harmonic oscillator $V(x) = -V_0 + \frac{V_0}{2a} x^2$.
In the complementary regime $a^2 V_0 \ll 1$, we get cancellation at leading order and
\begin{equation}
k_0^2 \gtrsim 0 (aV_0)^2 + 32 a^6 V_0^4 + \cdots
\label{shallow-limit}
\end{equation}
To investigate if this cancellation is specific to the form of the variational test function that was used, we also consider the normalized wavefunction $\psi_1 = \left( \beta' \right)^{1/4} e^{-\sqrt{\beta'}|x| }$, where the variational parameter $\beta'$ is chosen so that it is dimensionally consistent with the earlier choice in the definition of $\psi_0$. The variational principle gives
$$
\int \left( \psi_1'^2 + V(x) \psi_1^2 \right) dx =\beta' - V_0 \,\sqrt{2 \pi a^2 \beta' }  
  \, \text{erfc}\left(\sqrt{2 a^2 \beta'} \right) e^{2 a^2 \beta' } \geq -k_0^2 
$$
It is again clear that the optimal $\beta'$ is in the ``scaling form" $\beta' = a^{-2} F(a^2 V_0)$. Indeed this is immediate from dimensional considerations. Considering the regimes $a^2 V_0 \gg 1$ and $a^2 V_0 \ll 1$ separately we get
$$
a^2 \beta' \approx \begin{cases}  \frac{\pi}{2}(a^2V_0)^2  & a^2 V_0 \ll 1 \\  \frac{1}{2}\sqrt{a^2V_0}  & a^2 V_0 \gg 1 \end{cases}
$$
In the regime $a^2 V_0 \gg 1$ we get the variational bound
\begin{equation}
k_0^2 \gtrsim V_0 - \frac{\sqrt{V_0}}{a}, 
\end{equation}
which is consistent in terms of the scaling of the first correction to the leading order behavior, but suboptimal, in comparison with the bound in~\eqref{deep-limit}. In the complementary regime $a^2 V_0 \ll 1$, we get 
\begin{equation}
k_0^2 \gtrsim \frac{\pi}{2} (aV_0)^2 
\label{delta-limit}
\end{equation}
showing no cancellation at leading order and giving a result 
consistent with the limit $V(x) \to -\sqrt{2 \pi} aV_0 \delta(x)$. 

The 3d isotropic Gaussian well is given by the potential $V(x_1,x_2,x_3) = -V_0 \exp\left(-\frac{x_1^2+x_2^2 +x_3^2}{2a^2}\right)$.  Considering the product test function $\Psi_0(x_1,x_2,x_3) = \psi_0(x_1) \psi_0(x_2) \psi_0(x_3)$ we get the variational bound
$$
\int \left( |\nabla \Psi_0|^2 + V(x) \Psi_0^2 \right) d^3x = \frac{3\beta}{2} - V_0 \left(\frac{2 a^2 \beta}{1+2a^2\beta}\right)^{\frac{3}{2}} \geq -k_0^2 
$$
For sufficiently small $\beta$, the bound is asymptotically given by $\frac{3 \beta}{2} - \sqrt{8} V_0 a^3 \beta^{3/2}$, and is positive. Likewise, for large $\beta$, the bound is asymptotically equal to $\frac{3 \beta}{2} -  V_0$ which is also positive. This reflects the well known fact that attractive potentials in 3d do not support bound states, unless the potential is sufficiently deep.

To estimate the critical value of $V_0$ that allows for a bound state, we exploit the well known connection between the 3d and 1d Schr\"odinger equations \cite{griffithsQuantum}, that follows from the identity
$$
\frac{\partial^2 f}{\partial R^2} + \frac{2}{R} \frac{\partial f}{\partial R} = \frac{1}{R} \frac{\partial^2}{\partial R^2} (R f),
$$
so that for every spherically symmetric eigenfunction $\Psi(R)$ of the 3d operator $-\Delta + V(R)$ with a spherically symmetric potential, the function $\chi(R) = R \Psi(R)$ is an eigenfunction of the 1d operator $-\partial_{RR} + V(R)$ with the same energy. 

The converse, however, is not true. In order for the kinetic energy $ 4 \pi \int  |\nabla \Psi|^2 R^2 dR$ to be finite, we cannot have $\Psi(R)$ diverging as $1/R$ near $R = 0$. We therefore need that the corresponding 1d wavefunction $R \Psi(R)$ vanish at $R = 0$, i.e. we need a 1d bound state with a node at the origin in order to have a bound (ground) state for the 3d potential. The 1st excited state for the 1d gaussian well, which is the lowest energy state with a node at $R=0$, gives the energy of the ground state for the 3d gaussian well.

\begin{figure}[htbp] %  figure placement: here, top, bottom, or page
  \centering
  \includegraphics[width=0.8 \textwidth,trim={0cm, 1cm, 0cm, 2cm}, clip]{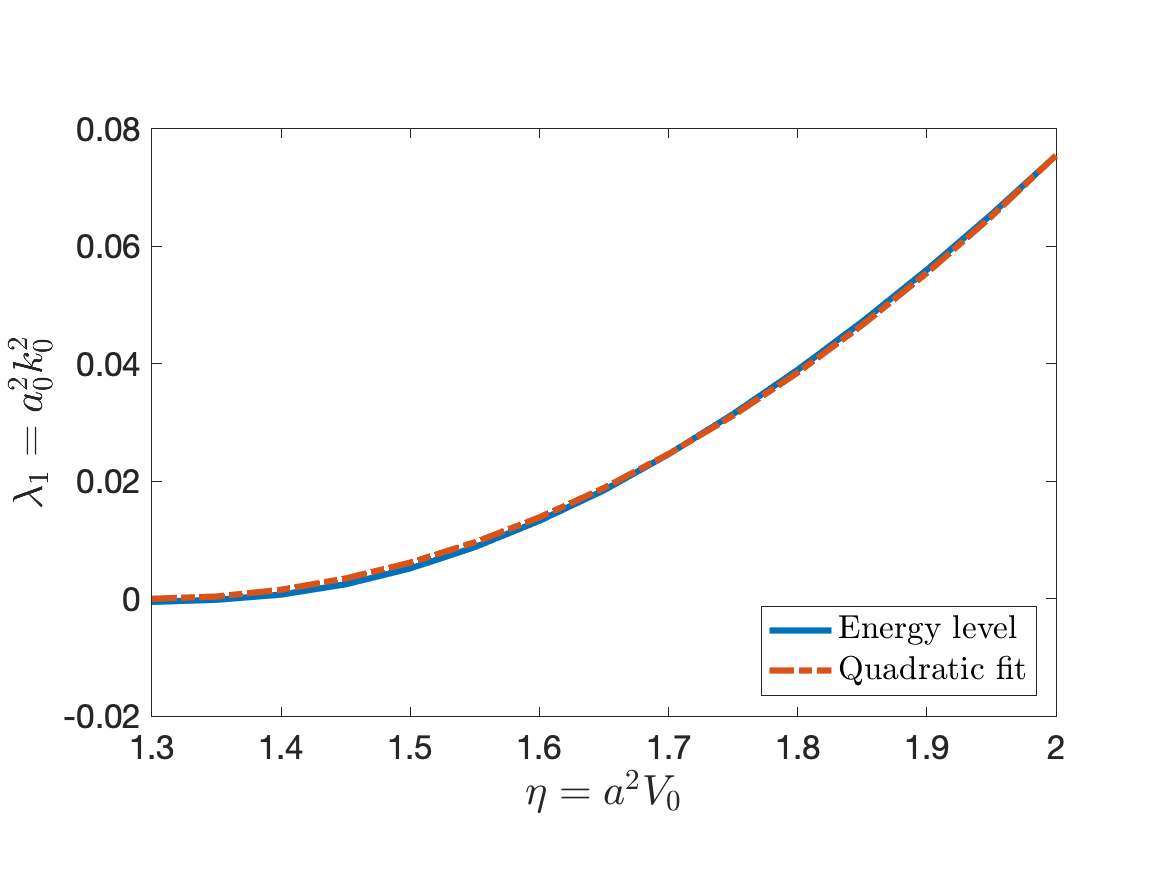} 
  \caption{The solid curve is energy level $\lambda_1$ of the first excited state of the 1d Gaussian potential $V(R) = \eta e^{-R^2/2}$. This state exists only if the well is sufficiently deep $\eta \geq \eta^* \approx 1.3$. A quadratic fit (the dashed curve) shows that $\lambda_1 \approx 0.154(\eta-1.3)^2$. }
  \label{bound-states}
\end{figure}

Fig.~\ref{bound-states} shows the numerically computed energy \cite{Driscoll2014} of the first excited state for the potential $W(R) = \eta e^{-R^2/2}$. There is a critical value $\eta^* \approx 1.3$ and an $O(1)$ constant $c \approx 0.154$ such that the first excited energy level satisfies
\begin{equation}
\lambda_1 = a^2 k_0^2 \approx c (V_0 a^2 - \eta^*)^2 .
\label{ball-k0}
\end{equation}
For applications to LSB disk galaxies, we also consider the anisotropic (oblate) potential $V(x_1,x_2,x_3) = -V_0 \exp\left(-\frac{x_1^2+x_2^2}{2a^2} - \frac{x_3^2}{2b^2}\right)$ with $b \ll a$. We now consider the product test function $\Psi =\psi_0(x_1) \psi_0(x_2) \psi_1(x_3)$ to allow for the possibility that $V_0 a^2 \gg1$ but $V_0 b^2 \ll 1$.  We now get the variational bound
$$
\int \left( |\nabla \Psi|^2 + V(x) \Psi^2 \right) d^3x = \beta+\beta' - \sqrt{2 \pi b^2 \beta' }  
  \, \text{erfc}\left(\sqrt{2 b^2 \beta'} \right) e^{2 b^2 \beta' } \frac{2 a^2 \beta V_0}{1+2a^2\beta} \geq -k_0^2 
$$
Since $b^2 V_0 \ll 1$ it is suggestive that $\beta' b^2 \ll 1$. We will assume this provisionally, and verify later that the assumption is valid. With this assumption, we have the variational bound
$$
k_0^2 \geq \frac{2 a^2 b \beta \sqrt{2 \pi \beta'} V_0}{1+2a^2\beta} - \beta - \beta'
$$
Optimizing over $\beta'$, we get that the optimal $\beta'$ is given by 
$$
b^2 \beta' = \frac{\pi}{2} \left( \frac{2 a^2 b \beta}{1+2a^2\beta}\right)^2 (b^2 V_0)^2
$$
so that our assumption that $b^2 \beta' \ll 1$ verifies. Also, we have the variational bound
\begin{equation}
k_0^2 \geq \beta  \left( \pi \frac{2 a^2 \beta}{(1+2a^2\beta)^2} \cdot (a^2 V_0) \cdot (b^2 V_0) -1\right) 
\label{disk-k0}
\end{equation}
From this bound, it follows that there exists a bound $\eta_c'$ such that we have bound states for all $a b V_0 \geq \eta_c'$. The optimal $\beta \sim a^{-2}$ and for $ab V_0 \gg 1$, we have $k_0 \sim b V_0$.

\section*{Acknowledgments}
SCV was partially supported by the Simons Foundation through award 524875.  SCV and ACN were also partially supported by the National Science Foundation through award GCR-2020915.


\bibliographystyle{spphys}

\bibliography{patternDM}

\end{document}